\documentclass[twocolumn]{aastex62}
\usepackage[utf8]{inputenc}

\def\msun{M_{\odot}}

\newcommand{\peryear}{yr$^{-1}$}

\shorttitle{X-rays from LISA's supermassive black holes}
\shortauthors{Dal Canton et al.}

\begin{document}

\title{Detectability of modulated X-rays from LISA's supermassive black hole mergers}

\correspondingauthor{Tito Dal Canton}
\email{tito@dalcanton.it}

\author[0000-0001-5078-9044]{Tito \surname{Dal Canton}}
\altaffiliation{NASA Postdoctoral Program Fellow}
\affiliation{NASA Goddard Space Flight Center, Greenbelt, MD 20771, USA}
\affiliation{LAL, Univ.~Paris-Sud, CNRS/IN2P3, Université Paris-Saclay, F-91898 Orsay, France}

\author[0000-0002-3689-1664]{Alberto Mangiagli}
\affiliation{Department of Physics G. Occhialini, University of Milano - Bicocca, Piazza della Scienza 3, 20126 Milano, Italy}
\affiliation{National Institute of Nuclear Physics INFN, Milano - Bicocca, Piazza della Scienza 3, 20126 Milano, Italy}

\author[0000-0003-3547-8306]{Scott C. Noble}
\altaffiliation{NASA Postdoctoral Program Fellow}
\affiliation{NASA Goddard Space Flight Center, Greenbelt, MD 20771, USA}
\affiliation{Department of Physics and Engineering Physics, The University of Tulsa, Tulsa, OK 74104, USA}

\author[0000-0002-2942-8399]{Jeremy Schnittman}
\affiliation{NASA Goddard Space Flight Center, Greenbelt, MD 20771, USA}

\author[0000-0001-5655-1440]{Andrew Ptak}
\affiliation{NASA Goddard Space Flight Center, Greenbelt, MD 20771, USA}

\author[0000-0001-5438-9152]{Antoine Klein}
\affiliation{Institut d’Astrophysique de Paris, CNRS \& Sorbonne Universités, UMR 7095, 98 bis bd Arago, 75014 Paris, France}
\affiliation{School of Physics and Astronomy and Institute of Gravitational Wave Astronomy, University of Birmingham, Edgbaston B15 2TT, UK}

\author[0000-0003-4961-1606]{Alberto Sesana}
\affiliation{School of Physics and Astronomy and Institute of Gravitational Wave Astronomy, University of Birmingham, Edgbaston B15 2TT, UK}

\author{Jordan Camp}
\affiliation{NASA Goddard Space Flight Center, Greenbelt, MD 20771, USA}

\date{\today}

\begin{abstract}
    One of the central goals of LISA is the detection of gravitational waves from the merger of supermassive black holes. Contrary to stellar-mass black hole mergers, such events are expected to be rich X-ray sources due to the accretion of material from the circumbinary disks onto the black holes. The orbital dynamics before merger is also expected to modulate the resulting X-ray emission via Doppler boosting in a quasi-periodic way, and in a simple phase relation with the gravitational wave from the inspiral of the black holes. Detecting the X-ray source would enable a precise and early localization of the binary, thus allowing many telescopes to observe the very moment of the merger. Although identifying the correct X-ray source in the relatively large LISA sky localization will be challenging due to the presence of many confounding point sources, the quasi-periodic modulation may aid in the identification. We explore the practical feasibility of such idea. We simulate populations of merging supermassive black holes, their detection with LISA and their X-ray lightcurves using a simple model. Taking the parameters of the X-ray Telescope on the proposed NASA Transient Astrophysics Probe, we then design and simulate an observation campaign that searches for the modulated X-ray source while LISA is still observing the inspiral of the black holes. Assuming a fiducial LISA detection rate of $10$ mergers per year at redshift closer than $3.5$, we expect a few detections of modulated X-ray counterparts over the nominal duration of the LISA mission.
\end{abstract}


\section{Introduction}

The Laser Interferometer Space Antenna (LISA) \citep{2017arXiv170200786A} is expected to observe many mergers of supermassive binary black holes, at rates between few and about a hundred per year \citep{Klein:2015hvg}. Such binary black holes have a high chance of residing in gas-rich environments, and accretion of the material on the black holes would produce copious amounts of radiation, including the X-ray band (e.g.~\citet{2002ApJ...567L...9A, 2012MNRAS.420..860S, 2016MNRAS.457..939C, 2018MNRAS.476.2249T, dAscoli:2018fjw}). The observation of such an electromagnetic counterpart to the LISA detection would enable the precise localization of the system on the sky, otherwise generally limited to a few arcminutes at best using the gravitational-wave data alone \citep{McWilliams:2011zs, Klein:2015hvg}. In addition, the joint observation of the gravitational wave and X-ray signal would provide valuable data on the physics of accretion and on the environment of black hole mergers. The identification of the correct electromagnetic source is nevertheless complicated by the weakness of the X-ray signal for distant sources, as well as by the presence of many additional confusing sources which are unrelated to the system emitting the gravitational waves. However, it has been suggested that the inspiraling orbital motion of the black holes and the accretion dynamics will produce a ``chirping'' modulation in the X-ray lightcurve, closely related to the gravitational-wave phase \citep{Haiman:2017szj}. It is then conceivable that the modulation could aid in the detection of the correct X-ray source among the myriad of confusing sources and X-ray background, once the imminent merger has been detected by LISA, and thus enable the precise localization of the source before the merger \citep{Kocsis:2007yu}. Similar ideas have also been proposed for LISA's white-dwarf binaries observed in the optical band \citep{Cooray:2003qm}.

Studying the electromagnetic counterparts to LISA sources is a major goal of several future missions involving electromagnetic observatories. One such mission, specifically designed to observe counterparts to gravitational-wave detections, is the NASA Transient Astrophysics Probe (TAP), a probe-scale mission proposed for the next decade \citep{2018AAS...23112105C}. Together with a set of instruments dedicated to studying kilonovae and gamma-ray-burst prompt and afterglow emissions, TAP includes an X-Ray Telescope (XRT) with 5 arcsec angular resolution, a relatively wide field of view of 1 deg$^2$ and 2 s timing precision. XRT is thus particularly well suited for the observation and precise localization of LISA's supermassive black hole mergers as described above. In addition, TAP is planned to have fast slewing capabilities and a halo orbit around the $L_2$ Lagrangian point, making most sky locations rapidly accessible and reducing the chance of blockage from the Sun, Earth and Moon to $\approx 15\%$ only.

In this paper we investigate the feasibility of this idea in detail. We perform simple simulations of X-ray-emitting supermassive black hole mergers and assume the currently planned characteristics of TAP/XRT and LISA. By simulating the entire process of LISA detection, tiling of the associated sky localization and X-ray observation with TAP/XRT, we determine the fraction of binaries detected by LISA which are also detectable in X-rays via their intrinsic modulation. We use a fiducial rate of LISA detections of supermassive black hole mergers to convert such fraction into expected rates of X-ray detections with TAP/XRT. This is done for different choices of the various parameters controlling the observation strategy, and we compare the choices in terms of associated detection rates and amount of required TAP observing time.

\section{Simulation setup}

We first describe how we simulate populations of supermassive black hole binaries and how we establish whether each binary leads to a detection of the X-ray modulation in addition to the gravitational-wave detection. We start with a broad overview of the process and give the details in the next subsections.

We first distribute many sources throughout the universe up to a maximum redshift, beyond which we do not expect any X-ray detection. We assign an X-ray lightcurve to each system based on its physical parameters. For each system we determine the time at which LISA detects its chirping gravitational-wave signal. At such time, we start observing the resulting sky localization probability density with our simulated X-ray telescope (XRT) using a simple algorithm for tiling the sky localization. We record the X-ray photons from the XRT pixel containing the true position of the source, and we search the photons for a modulation in their arrival times which is exactly in phase with the gravitational wave. If the search is successful, we record the earliest time at which the modulation is detected.

We can then repeat this process with different choices for the parameters of the population and the X-ray observation strategy, to study how the choices affect the fraction of successful detections and the distribution of detection time compared to the time of merger.

\subsection{Source population}

Our understanding of the population of merging supermassive black hole binaries is presently incomplete; unveiling its cosmic evolution will in fact be one of the major achievements of the LISA mission. Therefore, one assumption we need to make is the distribution of sources as a function of redshift and component masses. We base this assumption on the merger rate distributions presented in \citet{Klein:2015hvg}, which are approximately linear in redshift up to $z \approx 3$, $z \approx 5$ or $z \approx 9$ depending on the model. Because we do not expect an X-ray counterpart to be detectable at redshifts much larger than a few (and we will show later that this is indeed the case) here we adopt a linear distribution of sources in redshift, truncated at $z = 3.5$.

In \citet{Klein:2015hvg}, the distributions of total masses associated with different models are also discussed. Their models imply redshifted total masses that peak at values between $10^4 \msun$ and $10^7 \msun$, and considering that their detection rates peak between $z \approx 5$ and $z \approx 10$, the source-frame total mass $M$ would peak at $\approx 10$ times lower values. Because of hierarchical growth, however, the closest systems observed by LISA will also tend to be the most massive ones and the masses will still span the range $10^5 \lesssim M \lesssim 10^7$. Instead of choosing a particular scenario, we pick three representative populations of binaries with fixed masses: $(M,q) = \{ (5 \times 10^5 \msun, 1), (5 \times 10^6 \msun, 1), (5 \times 10^6 \msun, 10) \}$ where $q \ge 1$ is the ratio between the larger and smaller masses of the binary.

For simplicity, we assume zero spins on the black holes and no residual orbital eccentricity when the binaries become visible in LISA.

\subsection{Gravitational-wave observation}

Any sufficiently close inspiraling binary will affect the LISA data and produce a localized excess in the detection statistic used by the LISA search algorithm, assumed here to be the matched-filter signal-to-noise ratio (SNR). Initially the merger will be far in the future and the SNR induced in the LISA data will be undetectable. As LISA takes more and more data, however, the SNR will slowly increase, raise above the background and eventually lead to a confident detection of the binary at some time before its merger. For each source in our population, we thus calculate the SNR accumulated by LISA over time and we assume that a detection can be claimed as soon as the SNR reaches a threshold of 10. The SNR calculation is performed by using a gravitational waveform model which includes the inspiral, merger and ringdown epochs of the binary black hole evolution (IMRPhenomC, \citet{Santamaria:2010yb}).

After a detection, LISA will generate a posterior probability density for the sky localization of the source in the sky, similarly to what is routinely done by ground-based gravitational-wave detectors (e.g.~\citet{Singer:2015ema, GWTC1}). If the binary is detected sufficiently early, however, the localization will most likely receive periodic updates as more data is acquired, and will become more and more precise over time, ``zooming'' into the final estimate.

In order to simulate this process, we use a Fisher-matrix approach to obtain the size of the sky localization uncertainty as a function of time before merger. The Fisher-matrix calculation uses a frequency-domain waveform model which includes the inspiral part only, neglecting the merger and ringdown \citep{PhysRevD.90.124029}. Although this simplification would not be accurate for the SNR calculation, it is a sufficient approximation for the purpose of sky localization up to hours before merger. We relate the time and frequency of the inspiral using the post-Newtonian expression given in Eq.~2.7 of \citet{Lang:1900bz}.

In order to implement a realistic followup campaign with a limited field of view, we use the uncertainty area produced by the Fisher analysis to synthesize mock sky localization probabilities in the HEALPix format \citep{HEALPix} which is widely used in current gravitational-wave observations. This is done by drawing points from a 3-variate Gaussian distribution, projecting the points on the unit sphere, and binning them into a HEALPix array. The covariance of the Gaussian distribution is chosen so as to match the corresponding Fisher area, in the limit where the latter corresponds to a small fraction of the sky.

We assume that LISA will provide an updated sky localization every two days starting from the moment of first detection. This might sound conservative, but one has to consider the time needed to downlink the LISA data \citep{2017arXiv170200786A} and process it to obtain an updated sky localization. For the most interesting cases of detections well before the merger we do not expect a higher update cadence to make a large difference, since the sky localization will most likely be evolving slowly at that point.

The LISA sensitivity assumed in our calculation corresponds to the so-called ``ESACall v1.1'' configuration: the arm length is $2.5 \times 10^9$ m, the laser power is set to 2 W, the constellation uses six laser links and the results of LISA Pathfinder have been included. The same sensitivity is assumed, for instance, in \citet{PhysRevD.95.103012}.

\subsection{X-ray lightcurve}

\begin{figure}
    \includegraphics[width=\columnwidth]{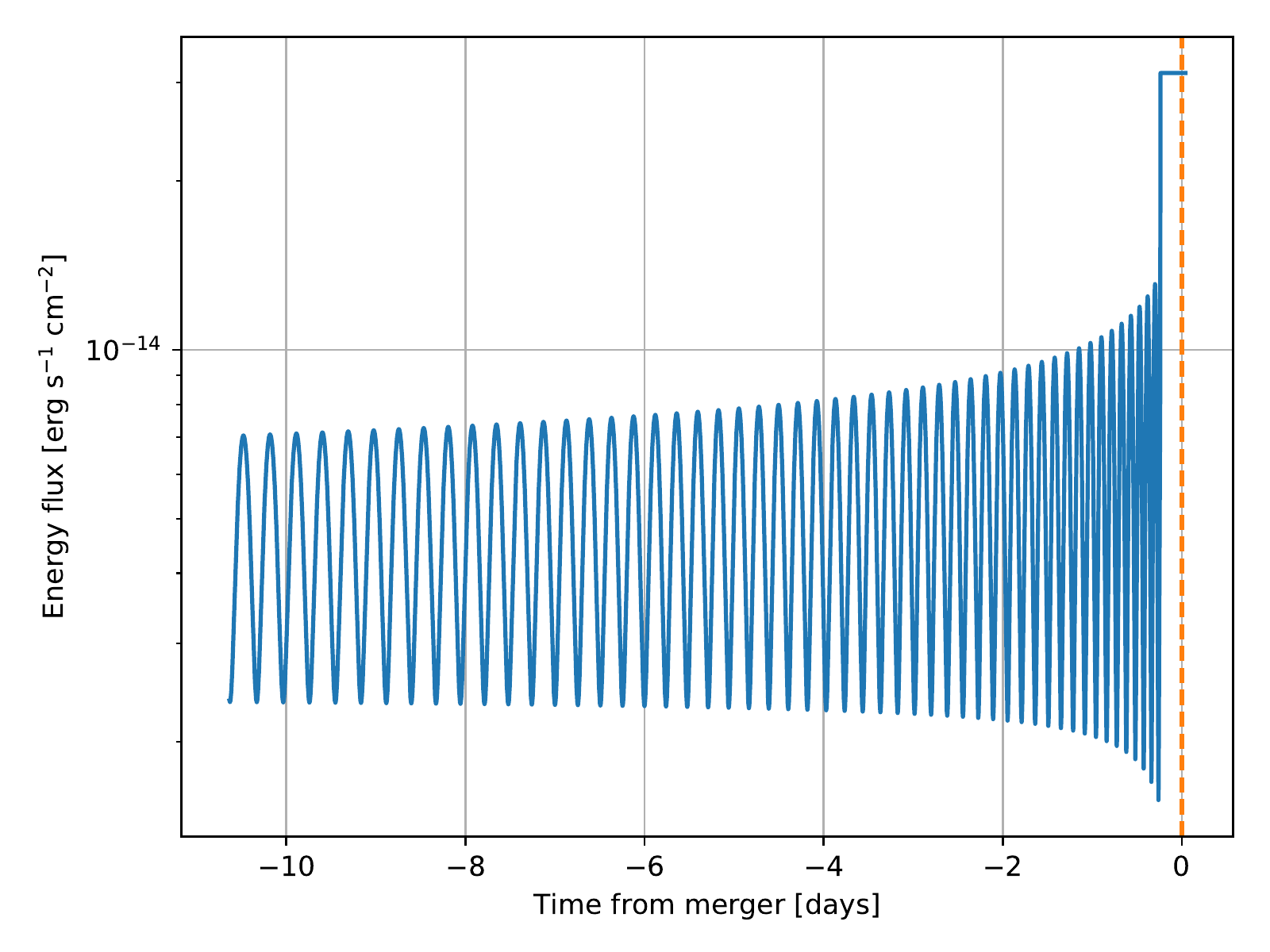}
    \caption{Example X-ray lightcurve for a binary with total mass $5 \times 10^6 \msun$, mass ratio 10, luminosity distance $13.6$ Gpc and inclination $40.3$ deg.}
    \label{fig:flux_lightcurve}
\end{figure}

The next component is a model of the X-ray emission. A complete physical model of the X-ray signal requires GRMHD simulations with radiation transport such as those described in \citet{dAscoli:2018fjw}. Currently available simulations only span a few orbits of the binary due to computational expense. Producing models that can cover days or months of observations is an open research problem and well outside the scope of this paper. Here we use a toy model with reasonable choices for how the flux might depend on the source parameters, with a computational cost enabling a large number of rapid simulations.
We model the X-ray flux at the solar system as
\begin{equation}
    F = \frac{\mu X L}{4\pi d^2_L}
    \label{eq:xrayflux}
\end{equation}
where $d_L$ is the luminosity distance and $L$ is the Eddington luminosity associated with the total mass of the binary. We assume the presence of minidisks around the black holes, and that one minidisk is much more luminous than the other. This configuration can be motivated based on the simulation described in \cite{Bowen2018}, which suggests that a non-axisymmetric overdensity in the circumbinary disk feeds preferentially one minidisk at each time. We do not include the emission from material that does not move with one of the black holes, such as the circumbinary disk. The factor
\begin{equation}
    \mu = 3 \sin(\iota) \cos(\psi) v + 1
\end{equation}
then represents the orbital modulation caused by Doppler beaming of the emission from the brighter minidisk, $\iota$ is the angle between the orbital angular momentum and the line of sight, $\psi$ is the orbital phase and $v$ the orbital velocity in natural units. The modulation is clipped such that it is always nonnegative, and it is shut down as soon as the binary reaches the frequency corresponding to the innermost stable circular orbit of a Schwarzschild black hole with mass $M$. The bolometric correction factor
\begin{equation}
     X = 0.1 + \frac{1}{a}
     \label{eq:xbolcor}
\end{equation}
converts the bolometric Eddington luminosity into the luminosity in X-rays and we assume it grows as the orbital separation $a$ decreases. The constant term $0.1$ is based on typical observations of active galactic nuclei \citep{Marconi2004,Lusso2012}, an X-ray spectrum with power-law index $-1.7$ typical of quasars \citep{Trakhtenbrot:2017xiz} and an XRT energy range of $0.5$-$2$ keV. The second term $a$, expressed in units of gravitational radii, is based on the following argument. Numerical simulations of merging black holes in a magnetized plasma show that the available magnetic energy increases towards merger with a roughly $a^{-1}$ scaling \citep{PhysRevD.96.123003}. This magnetic energy will likely be coupled to electron heating in the corona around each black hole mini-disk.
Another potential source of coronal heating is the collision of ballistic streams of gas from the circumbinary disk with the individual black hole mini-disks. This mechanism also naturally leads to a $a^{-1}$ scaling, as the outer edge of the mini-disks is proportional to $a$, so the potential energy drop across the gap will scale like $a^{-1}$ \citep{Roedig:2014cea}. The hot electrons in the corona will then efficiently produce X-rays by inverse Compton scattering the seed UV flux from the accretion disks around each black hole \citep{Schnittman2013, dAscoli:2018fjw}. Inverse Compton emission generically leads to a power-law spectrum with cut-off energy around 100 keV, so the X-ray flux in any given band should scale like the total inverse Compton power \citep{RybickiLightman}.
An example of the X-ray flux lightcurve predicted by our model is shown in Fig.~\ref{fig:flux_lightcurve}.

It should be noted that Doppler boosting is only one of the possible mechanisms leading to a quasi-periodic modulation of the X-ray lightcurve. The intrinsic variability of the accretion rate itself has also been proposed as a contributor to a quasi-periodic modulation \citep{Macfadyen:2006jx, 2012MNRAS.420..860S, Farris:2013uqa, DOrazio:2015shf, 2018MNRAS.476.2249T}, possibly even more important than Doppler boosting due to its different dependence on orbital inclination \citep{Kelley:2018fur}. Here we focus on Doppler boosting only, due to its relative model independence and certainty. It can be thought of as a lower limit on the estimated amplitude of X-ray variability in circumbinary systems.

\subsection{Sky localization tiling and photon measurement}

\begin{figure*}
    \includegraphics[width=0.32\linewidth]{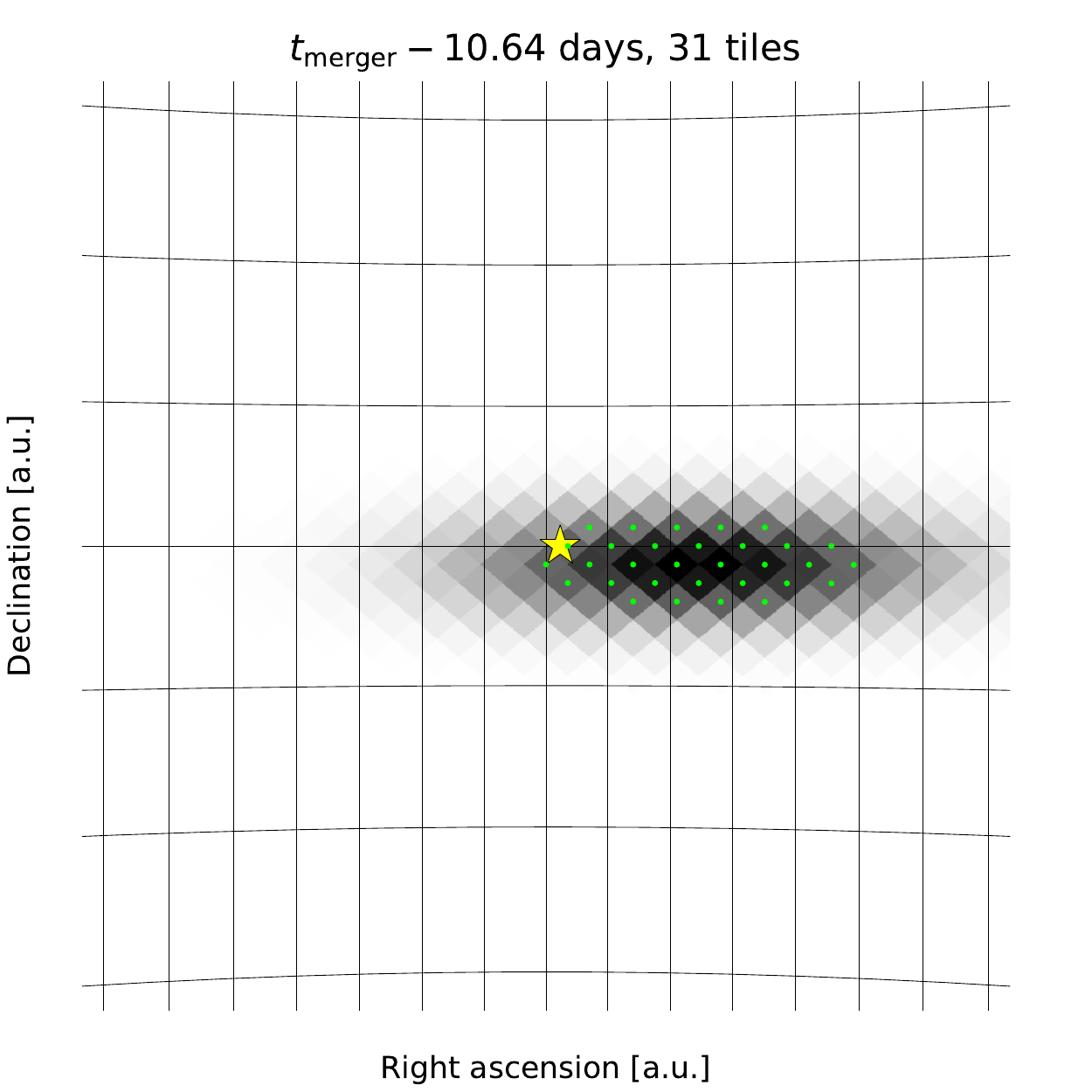}
    \includegraphics[width=0.32\linewidth]{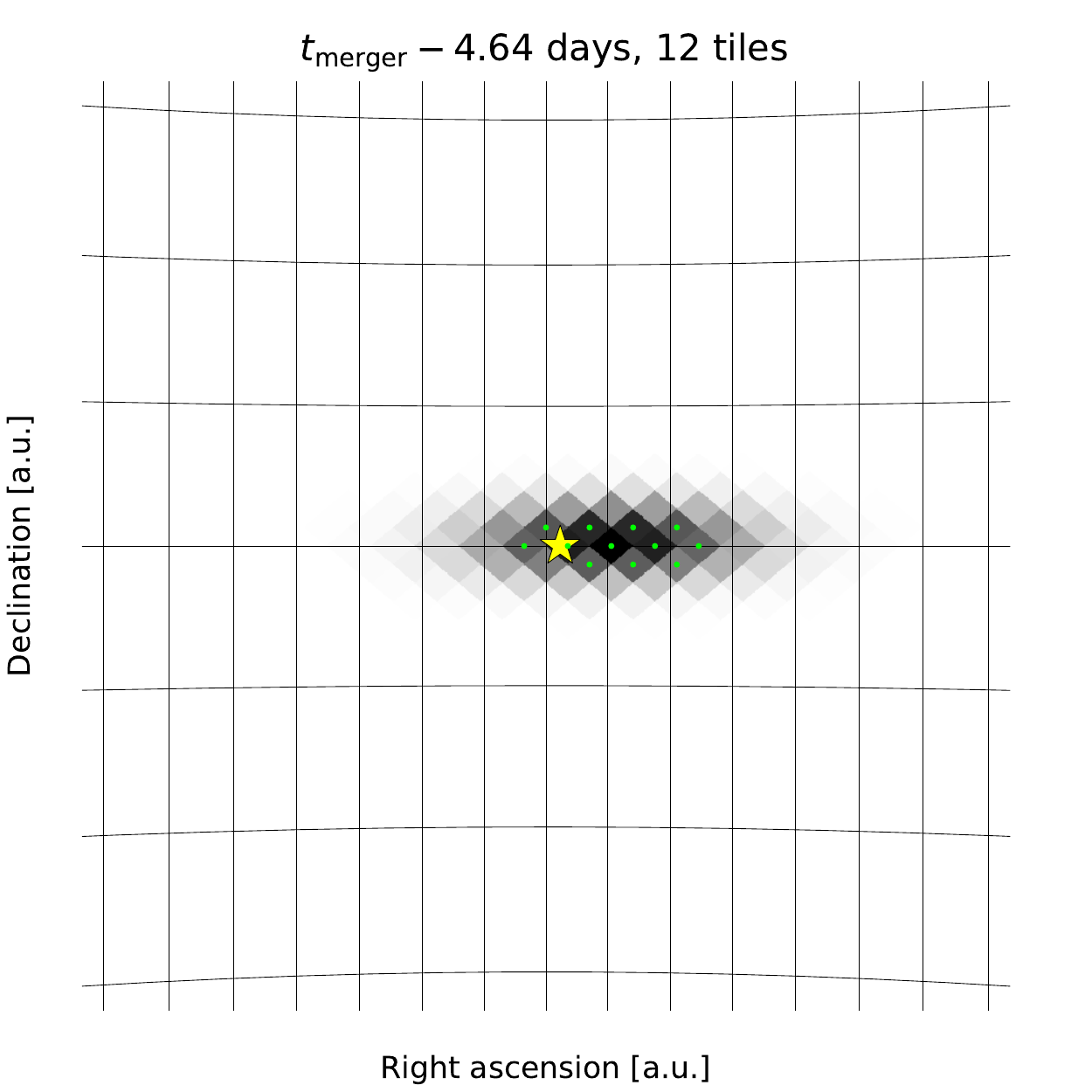}
    \includegraphics[width=0.32\linewidth]{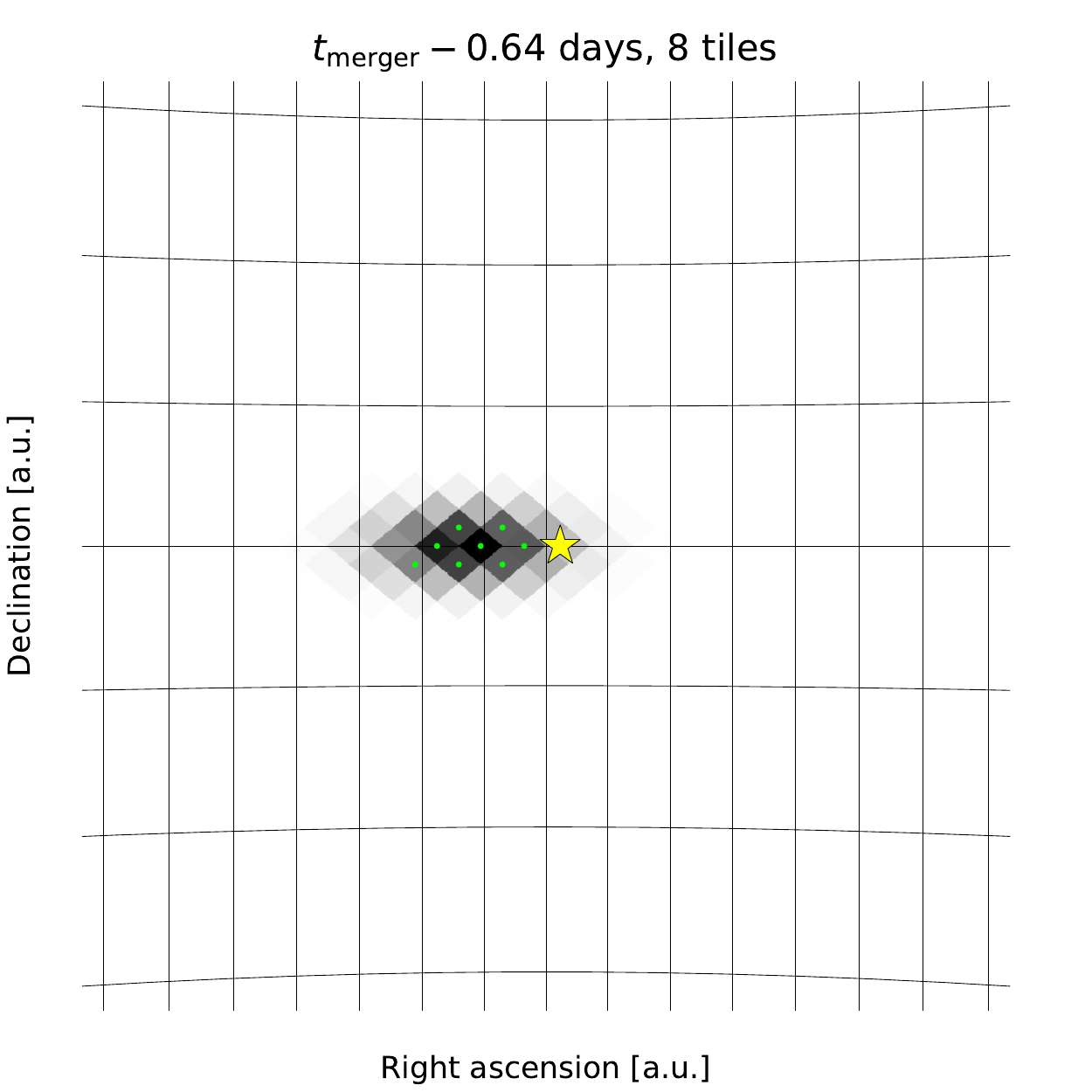}
    \caption{Example of the progression of the LISA sky localization and corresponding XRT tiling for the binary merger considered in Fig.~\ref{fig:flux_lightcurve}. Time and amount of considered gravitational-wave data both increase from left to right, and time is indicated above the plots. The central star is the true position of the source. The gray pixels show the XRT tiling of the LISA localization. Their darkness indicates the probability of containing the true source inferred from the LISA data, darker being more probable. The target coverage of the localization is set to 50\% and the tiles scheduled for XRT observation are indicated by a dot.}
    \label{fig:tiling}
\end{figure*}

We assume a square TAP/XRT field of view with 1 deg$^2$ area. The LISA sky localization probability map is tiled with XRT fields\footnote{Because we simulate the tiling using a HEALPix pixelization, the actual field of view is effectively a bit smaller, $0.84$ deg$^2$, which will make our results slightly conservative. Note that in an actual observation the fields will most likely have some overlap.} up to a fixed target covered probability. Each field is then observed in sequence in order of decreasing probability and with a constant exposure time until the target probability is reached. At that point the process starts again from the tile of highest probability. In addition, when LISA provides an updated sky localization, the current set of tiles expires and the next observation starts from the tile of maximum probability in the updated sky localization. The target probability coverage and the constant exposure time are parameters of our observation strategy which can in principle be optimized. We will explore the effect of different choices for these parameters when discussing the results later. An example of the evolution of the sky localization and its tiling is shown in Fig.~\ref{fig:tiling}.

We assume a TAP slew rate of 1 deg/s and a settle time of 10 s after each slew before starting the new exposure. We do not account for the time during which XRT or the source are blocked by the Sun, Earth or Moon, which amounts to a $\approx 15\%$ chance. For most of our simulations we also assume that TAP spends 100\% of its time observing each system, but we will eventually relax this as described later.

Using the current TAP/XRT design parameters, and a conservative energy band between $0.5$ and $2$ keV, we obtain a conversion factor from X-ray energy flux to photon rate at one XRT pixel of $1.26 \times 10^{12}$ cm$^2$ erg$^{-1}$. The resulting variable photon rate lightcurve is added to a constant rate of X-ray background photons ($7.2 \times 10^{-5}$ s$^{-1}$) and a constant rate of non-X-ray background ($2 \times 10^{-6}$ s$^{-1}$). The non-X-ray background rate is based on measurements in low-Earth orbit and is likely an underestimation of the background at $L_2$; however, it would have to be significantly larger to compete with the X-ray background. The total photon rate time series is then used to draw a set of random photons via inverse transform sampling.

\begin{figure}
    \includegraphics[width=\columnwidth]{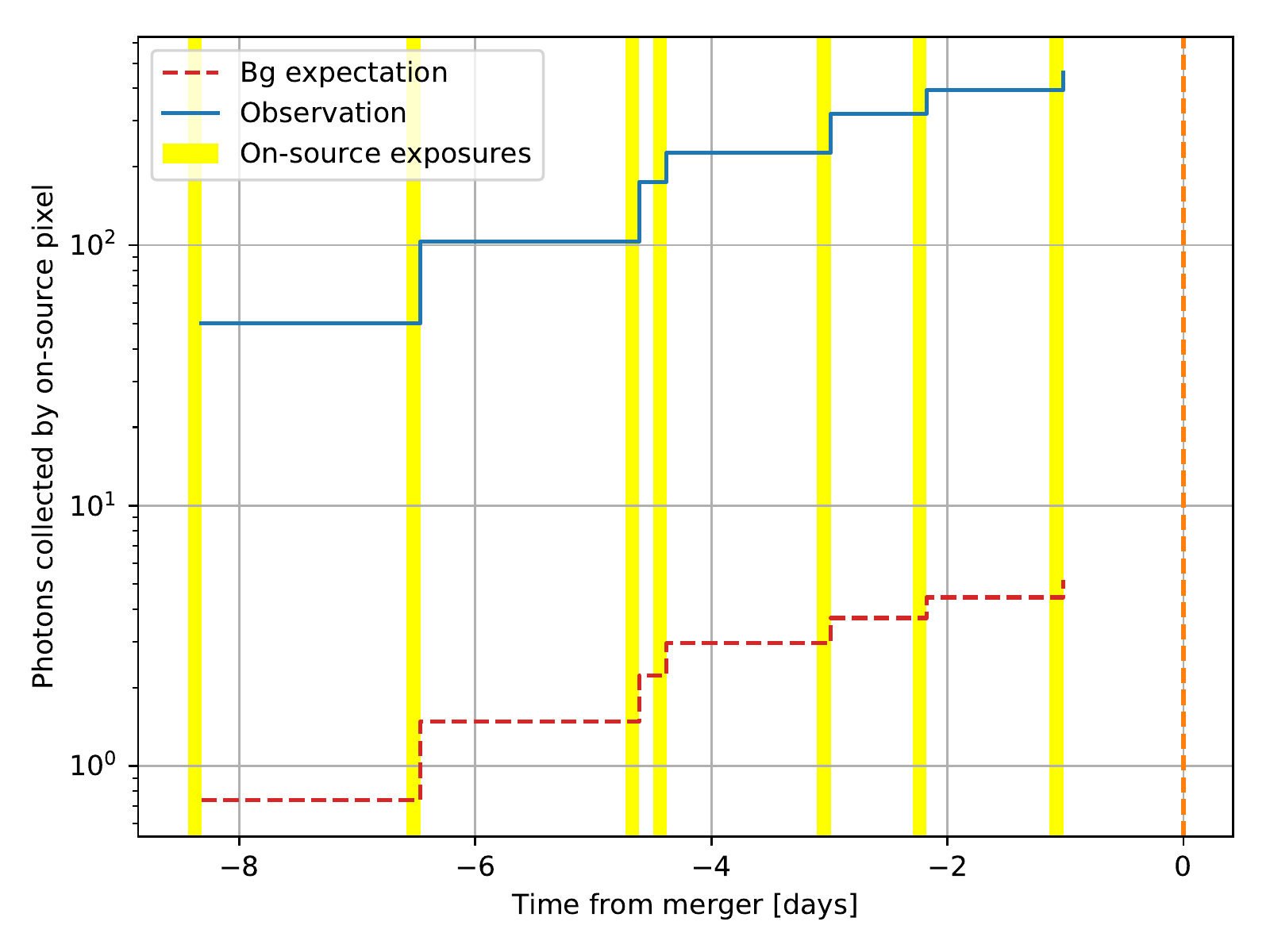}
    \caption{Number of photons collected over time by TAP/XRT at the pixel containing a simulated source with the lightcurve shown in Fig.~\ref{fig:flux_lightcurve}. The dashed steps show the number of photons expected in the absence of a source. At a luminosity distance of $13.6$ Gpc, this particular system is immediately detectable as a bright point source.}
    \label{fig:photon_accumulation}
\end{figure}

Fig.~\ref{fig:photon_accumulation} shows the number of photons accumulated over time for the binary simulated in Fig.~\ref{fig:flux_lightcurve}.
In this particular case, the system stands well above the background as a bright point source after a single exposure. However, many other point sources will be detected in the large LISA sky localization uncertainty, and there would be no way of recognizing that this is the correct source without an additional signature.

\subsection{Search for lightcurve modulation with uneven observations}

We now need a way to analyze the collected photons and search for a pulsation whose phase is related to the gravitational-wave phase. Various methods have been proposed and used for detecting modulations in X-ray lightcurves. We are looking for a method which (i) can directly use the photon arrival times without ad-hoc binning, (ii) can handle arbitrarily small number of photons, and (iii) can work with a photon sample produced by many irregular observations of the source, possibly partially synchronized or beating with the modulation itself.

If we had a complete and accurate model of the X-ray signal, similarly to what happens in searches for gravitational waves from compact binary mergers, a straightforward analysis fulfulling all of the above requirements would simply consist in computing the likelihood for observing the given set of photons under that model. However, it is well known that typical lightcurves from active galactic nuclei (AGN) exhibit stochastic behavior, and the accretion rate could have its own intrinsic variation independent of the modulation produced by the binary (e.g.~\citet{Mushotzky1993, Ulrich1997}). Therefore, we prefer not to assume that both the amplitude and phase of the signal will have very accurate models with low computational cost soon. The method we are looking for should focus instead purely on the periodicity, which is easier to model given the known dynamics of the inspiral, without taking the amplitude of the signal into account.

A simple frequentist method which fulfills all of the above requirements is Kuiper's test \citep{Paltani:2004wg}. Given a set of photon arrival times $\{t_i\}$, we can assign each photon a phase $\psi_i = \psi(t_i)$ by using the phase evolution $\psi(t)$ inferred from the gravitational-wave observation. The photon phases should be understood as modulus $2\pi$. Under the null hypothesis (absence of a modulation at the assumed phase), $\{\psi_i\}$ will be drawn from a distribution which purely reflects the density of observations at each point in the phase cycle. Under the presence of a modulation, instead, the underlying distribution will be distorted by the phase structure of the modulation (which we are not concerned with here). Calling $T(\psi)$ the empirical cumulative distribution function of $\{\psi_i\}$, and $U(\psi)$ the cumulative distribution of the observations associated with the null hypothesis, Kuiper's statistic is then
\begin{equation}
    \mathcal{K} = \max_{\psi}[T(\psi) - U(\psi)] + \max_{\psi}[U(\psi) - T(\psi)].
\end{equation}
The null distribution for $n_{\rm obs}$ observations can be calculated exactly as
\begin{equation}
    U(\psi) = \frac{1}{n_{\rm obs}} \sum^{n_{\rm obs}}_{n=1} \frac{\psi c_{n} + a_n(\psi)}{2\pi c_n + b_n(\psi)}
\end{equation}
where $c_n$ is the integer number of full phase cycles covered by observation $n$ and $a_n$, $b_n$ take into account whether or not $\psi$ falls within the remaining fraction of cycle in observation $n$.

Here we assume that the phase of the modulation is known \emph{exactly} as soon as LISA detects the gravitational-wave signal. After each exposure containing the source, we use the known phase to perform Kuiper's test on all photons accumulated so far by the on-source XRT pixel. We convert the resulting $\mathcal{K}$ to a $p$-value ($p_{\rm kuiper}$) by using the formulae from \citet{Stephens1965} and \citet{Paltani:2004wg} which are implemented in Astropy \citep{2013A&A...558A..33A,2018AJ....156..123A}\footnote{Note that the Astropy implementation of Kuiper's $p$-value prior to version 3.1 is incorrect, and that Eq.~7 of \citet{Paltani:2004wg} is incorrectly quoted from \citet{Stephens1965}.}.

In a real observation we will not know a priori which pixel contains the source and we will have to perform Kuiper's test on every pixel of every pointing, hence the rate of false positives will be inflated by a large trials factor $n_{\rm trials}$. We take $n_{\rm trials} = n_{\rm exp} n_{\rm pix}$, where $n_{\rm exp}$ is the total number of exposures that have been done up to a certain time, and $n_{\rm pix} = 518400$ is the number of pixels contained in the XRT field of view assuming the current design. Because some exposures might only add zero or a few photons to an otherwise already large sample, $n_{\rm exp}$ is most likely an overestimate; the exact value is likely somewhere between the number of different sky positions observed by XRT and $n_{\rm exp}$. We then apply the trials factor to Kuiper's $p$-value from the on-source pixel, producing a final $p$-value
\begin{equation}
    p^{\rm trials}_{\rm kuiper} = 1 - (1 - p_{\rm kuiper})^{n_{\rm trials}}.
    \label{eq:ptrials}
\end{equation}

\begin{figure}
    \includegraphics[width=\columnwidth]{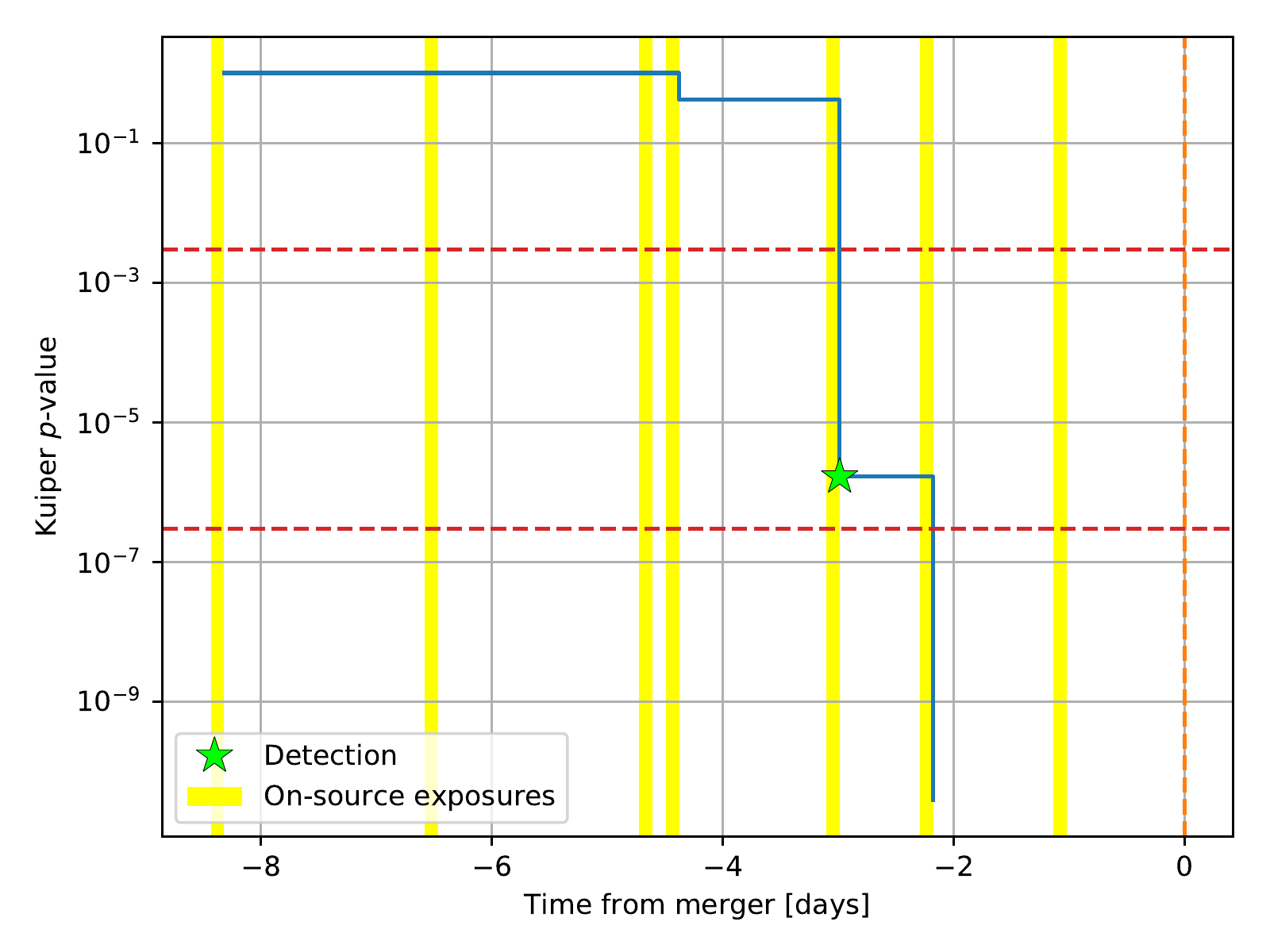}
    \caption{Evolution of the significance of Kuiper's test over time for the binary from Fig.~\ref{fig:flux_lightcurve}. The solid steps show the $p$-value of Kuiper's test for the XRT pixel containing the system. The many off-source curves are not shown, but the corresponding trials factor is taken into account in the solid curve. The dashed horizontal lines mark the $3\sigma$ and $5\sigma$ significance levels. After several exposures observing the system, the set of collected photons is large enough and covers a sufficient portion of the X-ray modulation cycle so as to make the modulation evident.}
    \label{fig:kuiper_over_time}
\end{figure}

We claim a detection of the X-ray modulation as soon as $p^{\rm trials}_{\rm kuiper}$ drops below a threshold of $0.003$, roughly corresponding to $3\sigma$ significance. Although this threshold is too low for a confident detection using TAP/XRT alone, we argue that many other instruments will likely want to observe the identified location as soon as $\sim 3\sigma$ is reached, and we assume they would either confirm or rule out the source within a short time. Note that the TAP/XRT observation continues until the binary has merged, regardless of whether a detection has been claimed or not. An example of the evolution of $p^{\rm trials}_{\rm kuiper}$ as more and more observations are made is shown in Fig.~\ref{fig:kuiper_over_time}.

Most of our simulations do not include the presence of bright confusing sources in addition to the true emitting system. Rather, we make the assumption that any other system within the LISA localization uncertainty is extremely unlikely to produce a lightcurve in phase with the true system so as to trick Kuiper's test into producing a false positive. Any confusing source would then effectively behave as the background noise, which we account for via the trials factor described above. We revisit this assumption and explore its validity in a later section.

\section{Results and discussion}

After having described the details of how the simulations are performed, in this section we explore and discuss the results for different choices of the various parameters that define the observation strategy.

\subsection{Baseline}

\begin{figure*}
    \includegraphics[width=0.666\columnwidth]{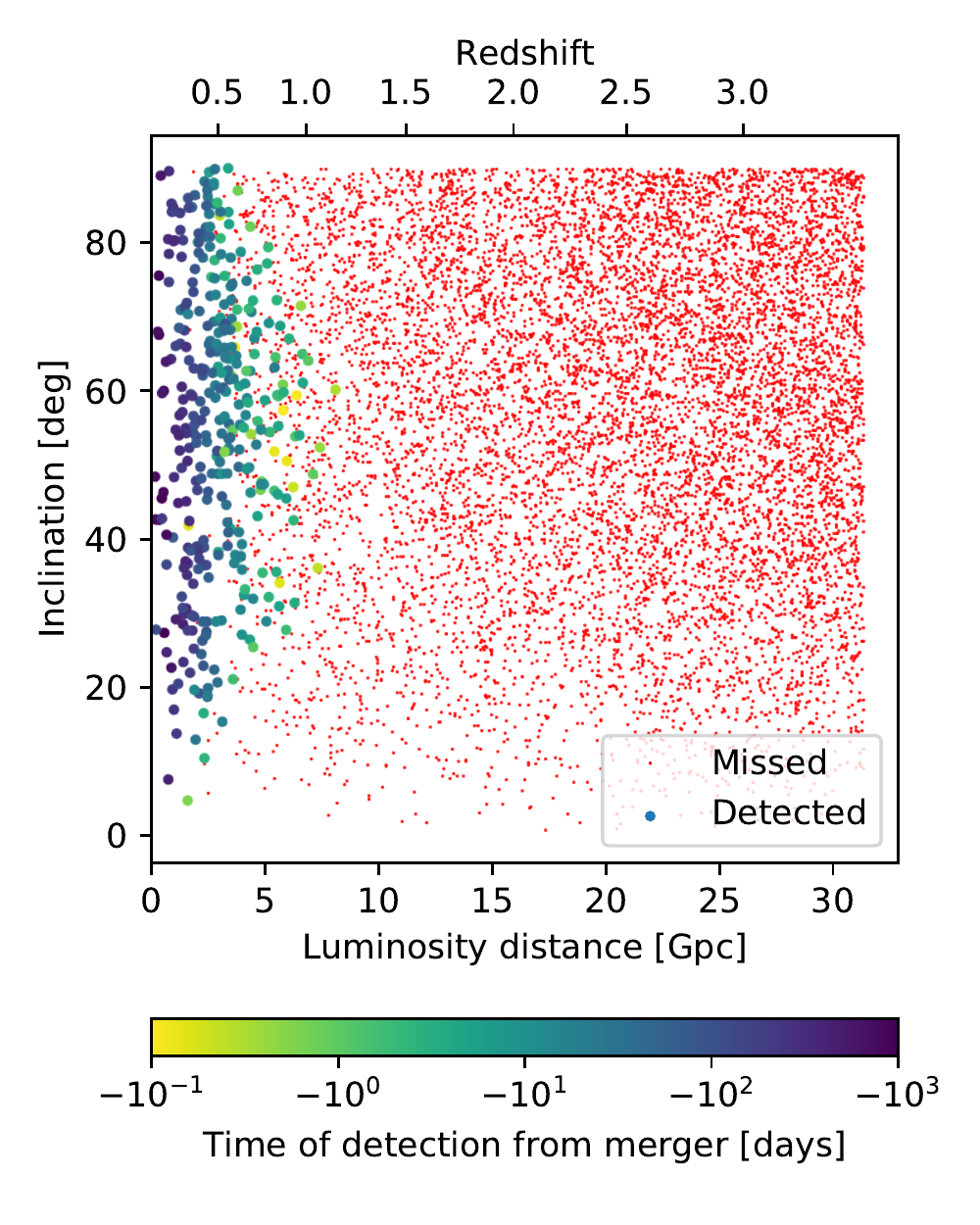}
    \includegraphics[width=0.666\columnwidth]{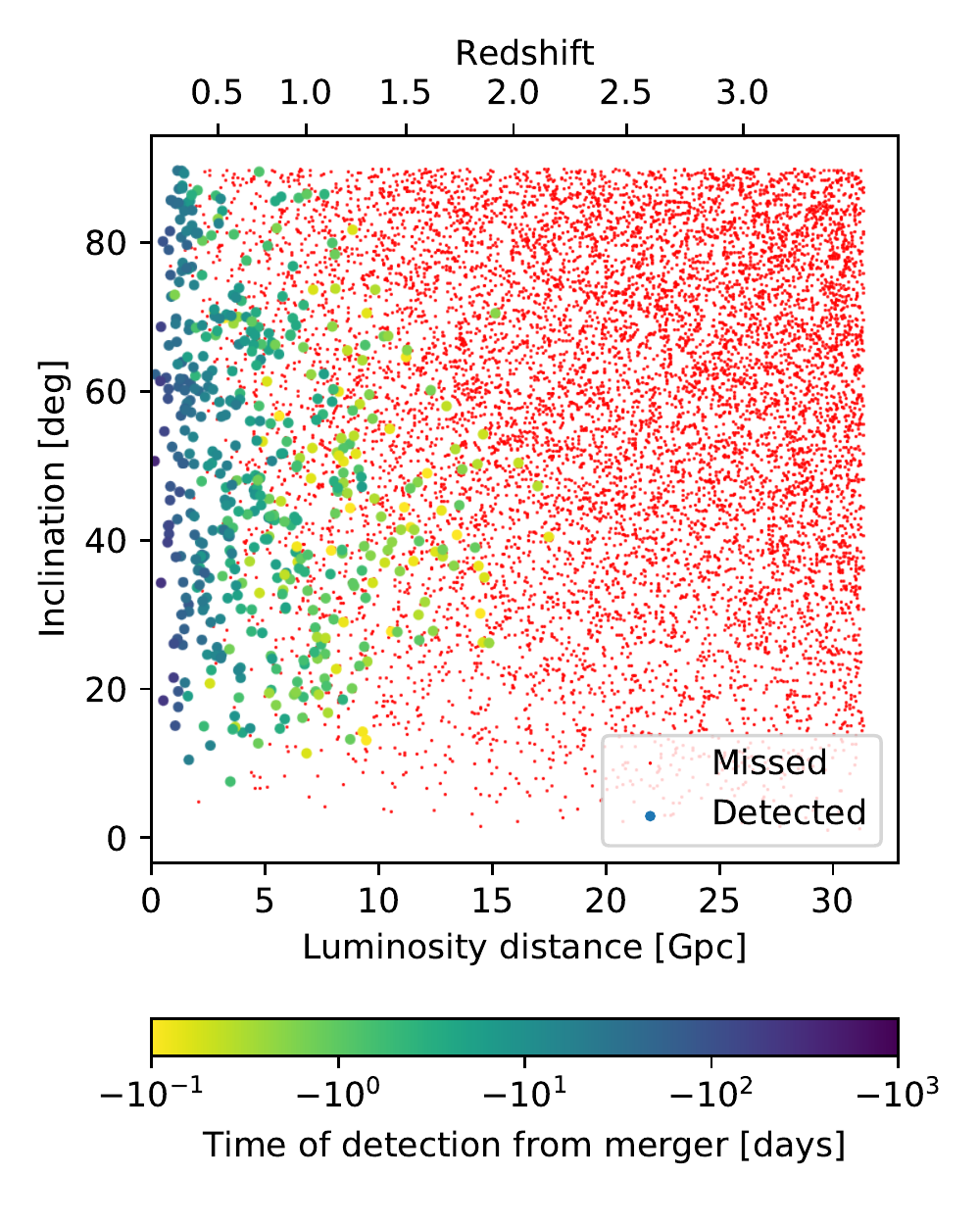}
    \includegraphics[width=0.666\columnwidth]{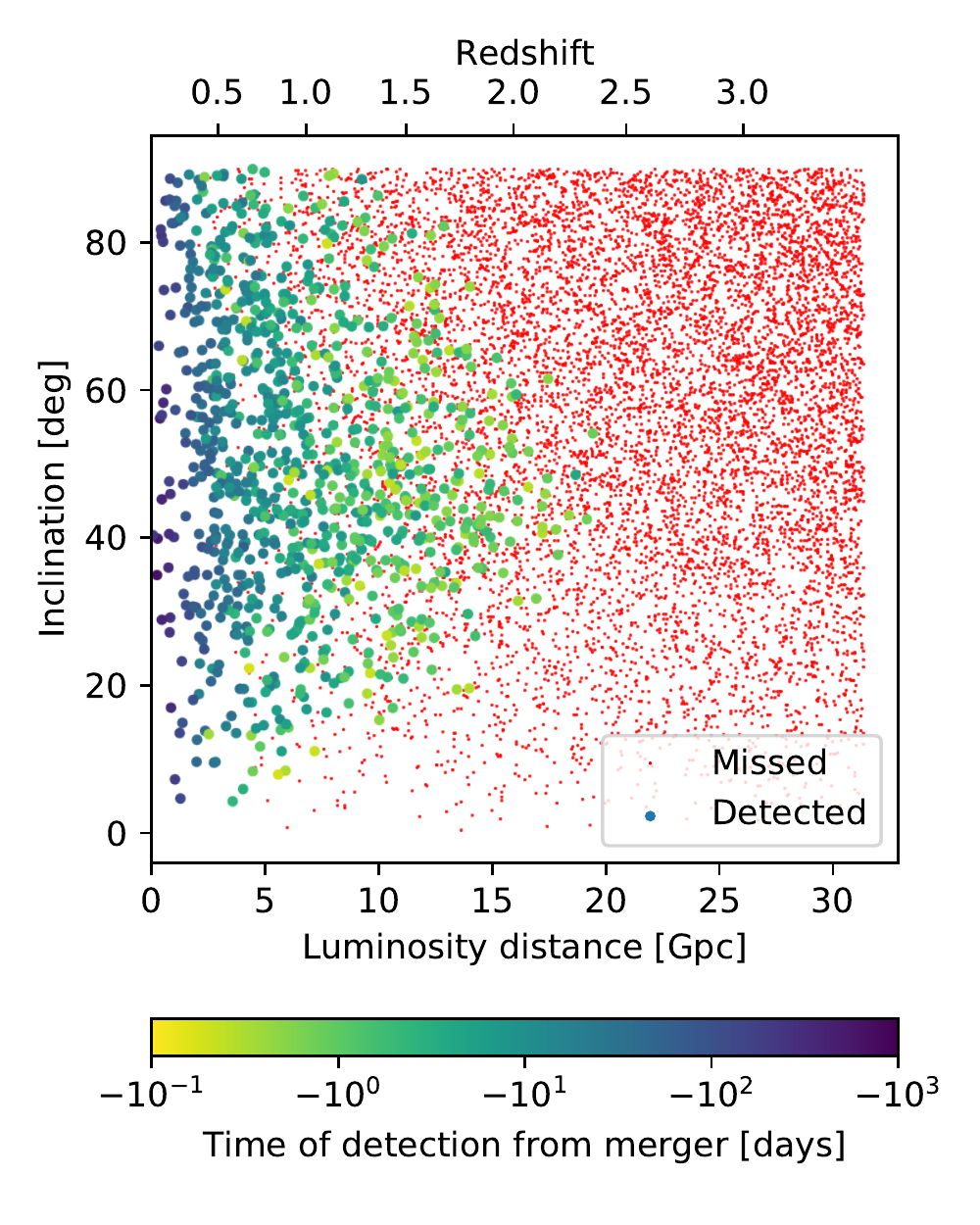}
    \caption{Distribution over distance and inclination of sources whose X-ray modulation was detected (larger dots) and missed (smaller dots). The color of the detected systems indicates how early before merger the detection happened, darker being earlier. From left to right: $M=5\times 10^5\ \msun$, $q=1$; $M=5\times 10^6\ \msun$, $q=1$; $M=5\times 10^6\msun$, $q=10$. Nearby systems are detected much earlier and over a broader range of inclinations than more distant ones. Higher total masses and higher mass ratios both increase the chance of detecting a system.}
    \label{fig:time_inc_dist}
\end{figure*}

We start the exploration by assuming a fixed XRT exposure of $10^4$ s per tile and a tiling strategy that aims at covering 50\% of the probability of the LISA localization uncertainty. Fig.~\ref{fig:time_inc_dist} shows the detected and missed systems ($10^4$ in total) as a function of their distance and orbital inclination with the line of sight, for the three choices of mass parameters described earlier: two equal-mass binaries of total mass $5\times 10^5 \msun$ and $5\times 10^6 \msun$, and an asymmetric binary of total mass $5\times 10^6 \msun$ and mass ratio 10.
The X-ray modulation from systems at $z \lesssim 0.5$ is detected tens or hundreds of days before merger. As we go to more distant systems, the times of detection are pushed closer and closer to merger, and detecting the modulation at all becomes very unlikely beyond $z = 2.5$. The latter observation validates our initial choice of truncating the redshift distribution at $3.5$.

The component masses have a strong impact on the detectability of the system: heavier systems are detectable to higher distances than lighter ones, and asymmetric systems are easier to detect than equal-mass ones. Systems with intermediate orbital inclinations ($\approx 45$ deg) also appear to be easier to detect than face-on and edge-on systems. This is not surprising, since face-on orientations suppress the Doppler modulation, while edge-on orientations suppress one polarization of the gravitational wave and are more difficult to localize with LISA.

\begin{figure*}
    \includegraphics[width=0.666\columnwidth]{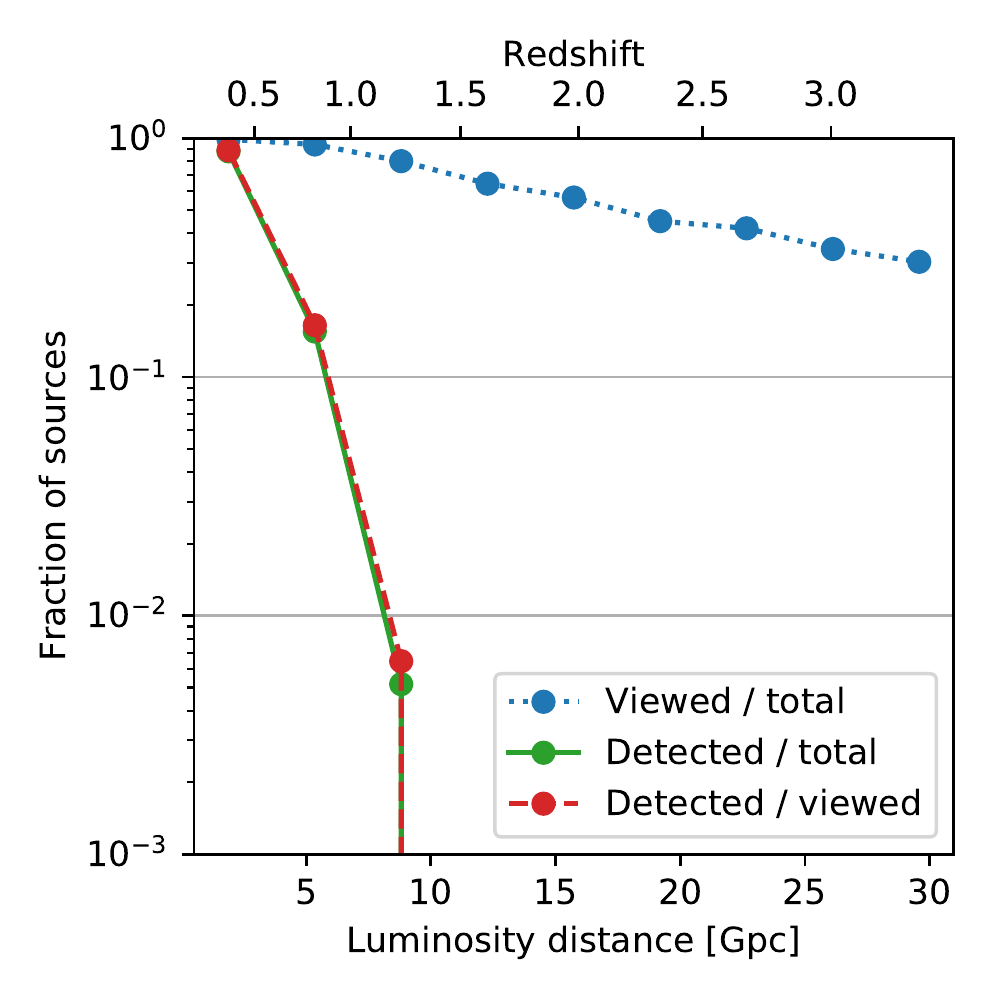}
    \includegraphics[width=0.666\columnwidth]{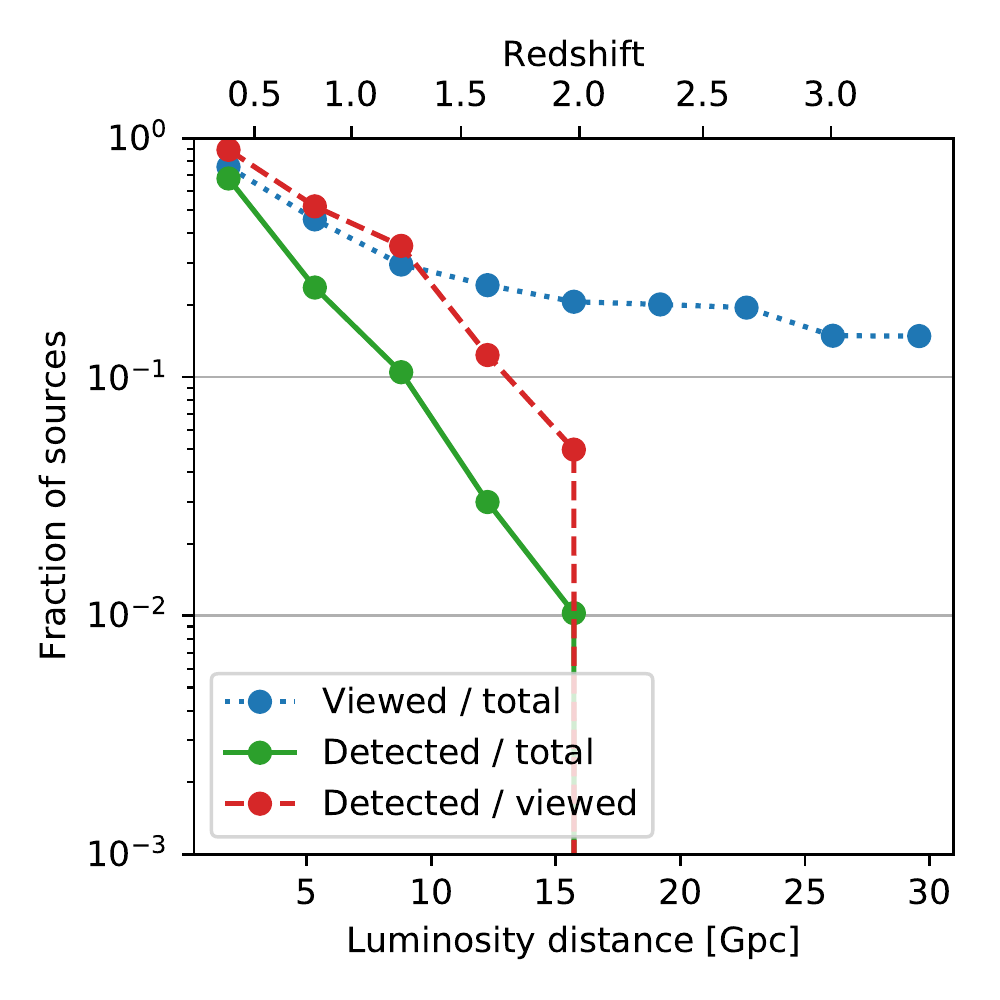}
    \includegraphics[width=0.666\columnwidth]{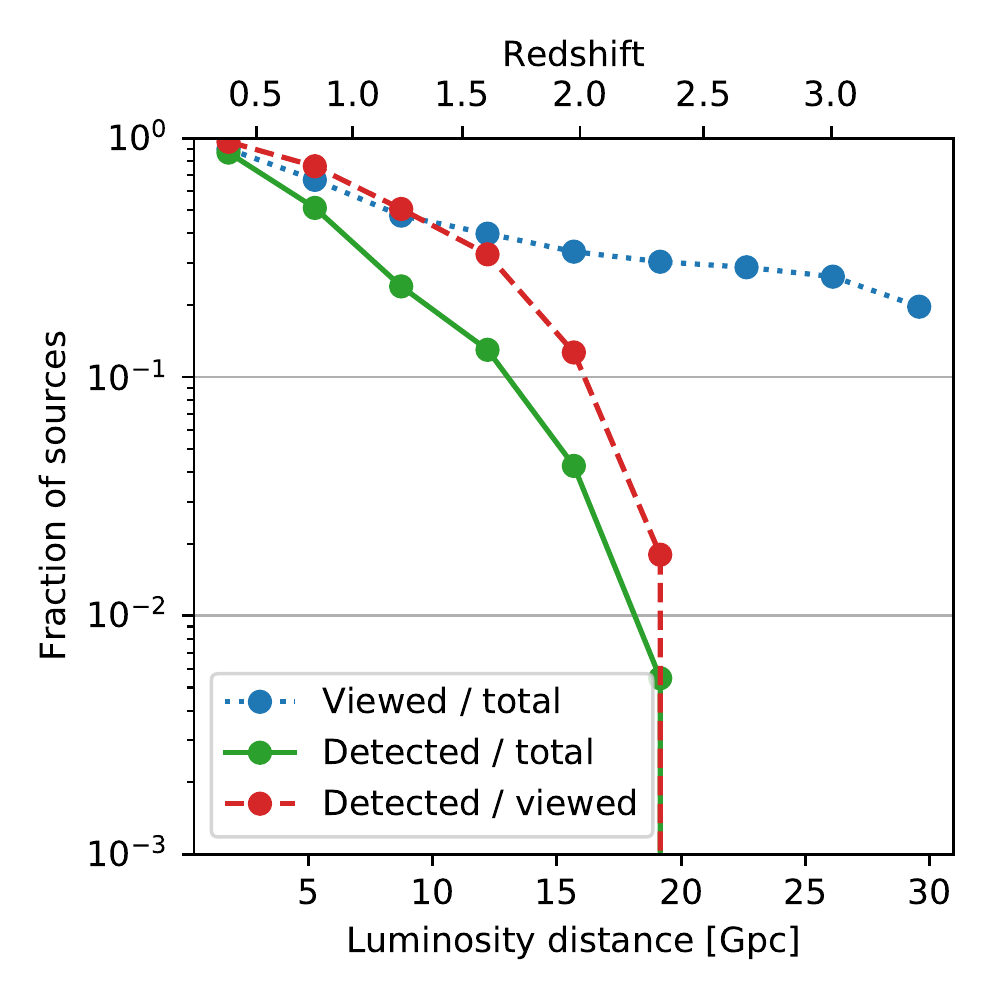}
    \caption{Fraction of viewed and detected sources in equal bins of luminosity distance. By ``viewed'' we mean sources that fall in the XRT field of view at least once, but may or may not produce a detection of their X-ray modulation. From left to right: $M=5\times 10^5\ \msun$, $q=1$; $M=5\times 10^6\ \msun$, $q=1$; $M=5\times 10^6\msun$, $q=10$.}
    \label{fig:fraction_dist}
\end{figure*}

A system can be missed for two reasons: (i) XRT never actually views its true location, for instance because the sky localization remains broad until the very end of the inspiral and the target coverage is limited; (ii) XRT views the system at least once, but the collected photons are insufficient for Kuiper's test to detect the modulation. The fractions of viewed and detected systems both decrease with distance, as shown in Fig.~\ref{fig:fraction_dist}, because the sky localization becomes broader and at the same time the X-ray flux decreases. The relative contribution of these two effects depends on the mass parameters. Lighter systems have a larger chance of being viewed at larger distances than heavier systems. This can be attributed to the fact that lighter systems are generally better localized by LISA, because their gravitational waveforms contain more signal power at high frequency and because they are modulated by the LISA orbital motion when they are already in the LISA frequency band. We also see, however, that the fraction of viewed systems leading to a detection drops much faster with distance for lighter systems than heavier ones.

\begin{figure*}
    \includegraphics[width=0.66\columnwidth]{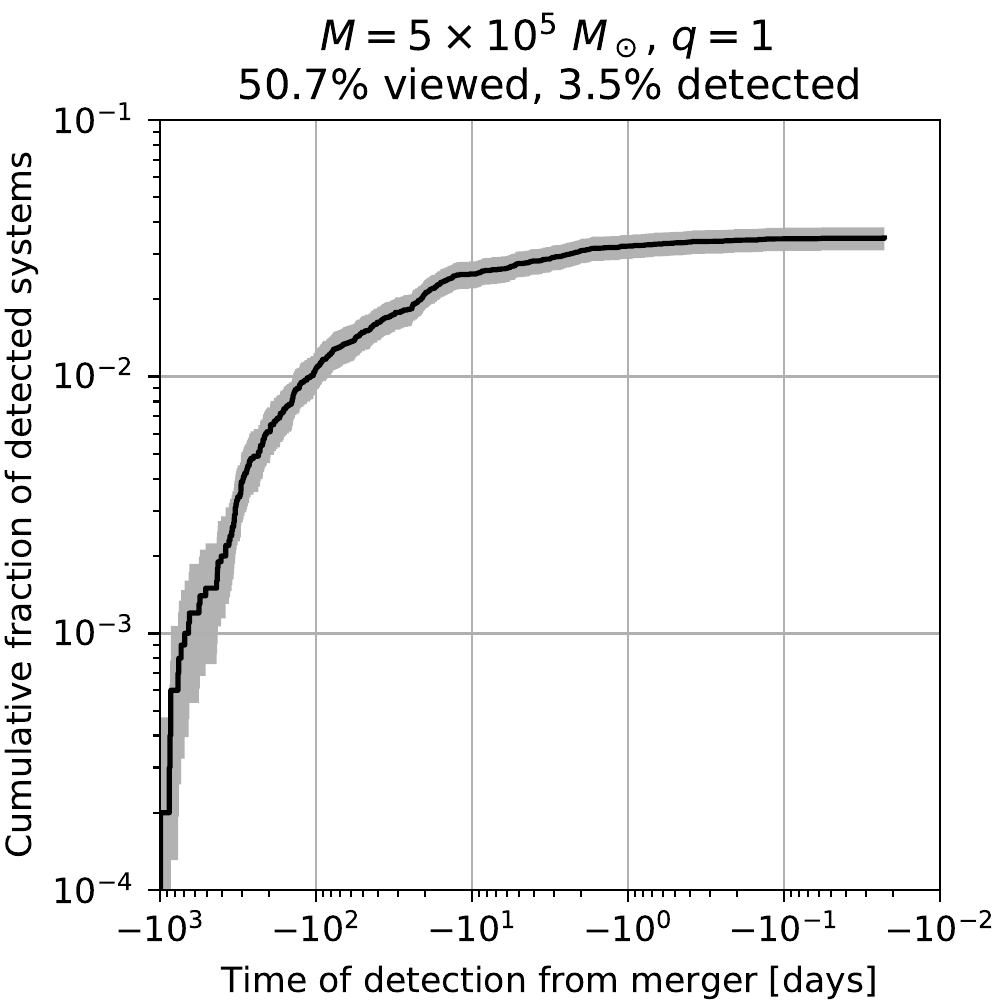}
    \includegraphics[width=0.66\columnwidth]{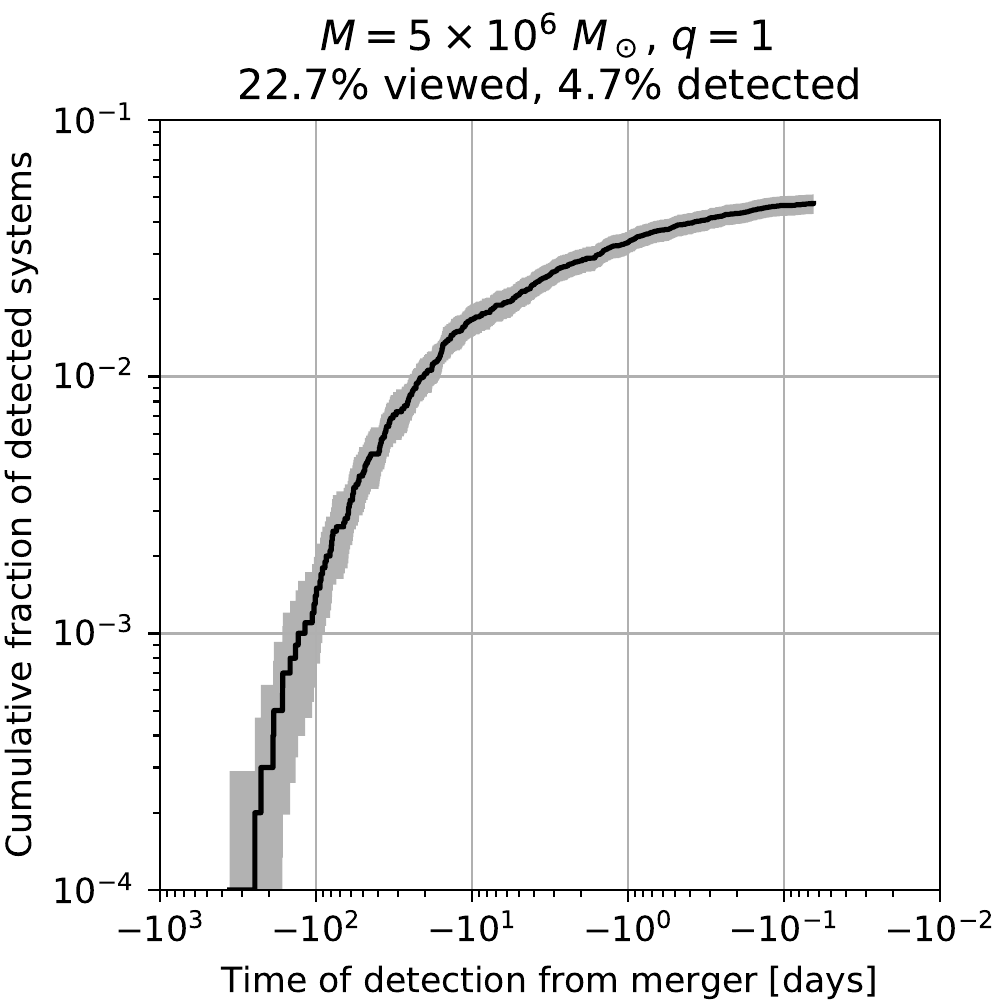}
    \includegraphics[width=0.66\columnwidth]{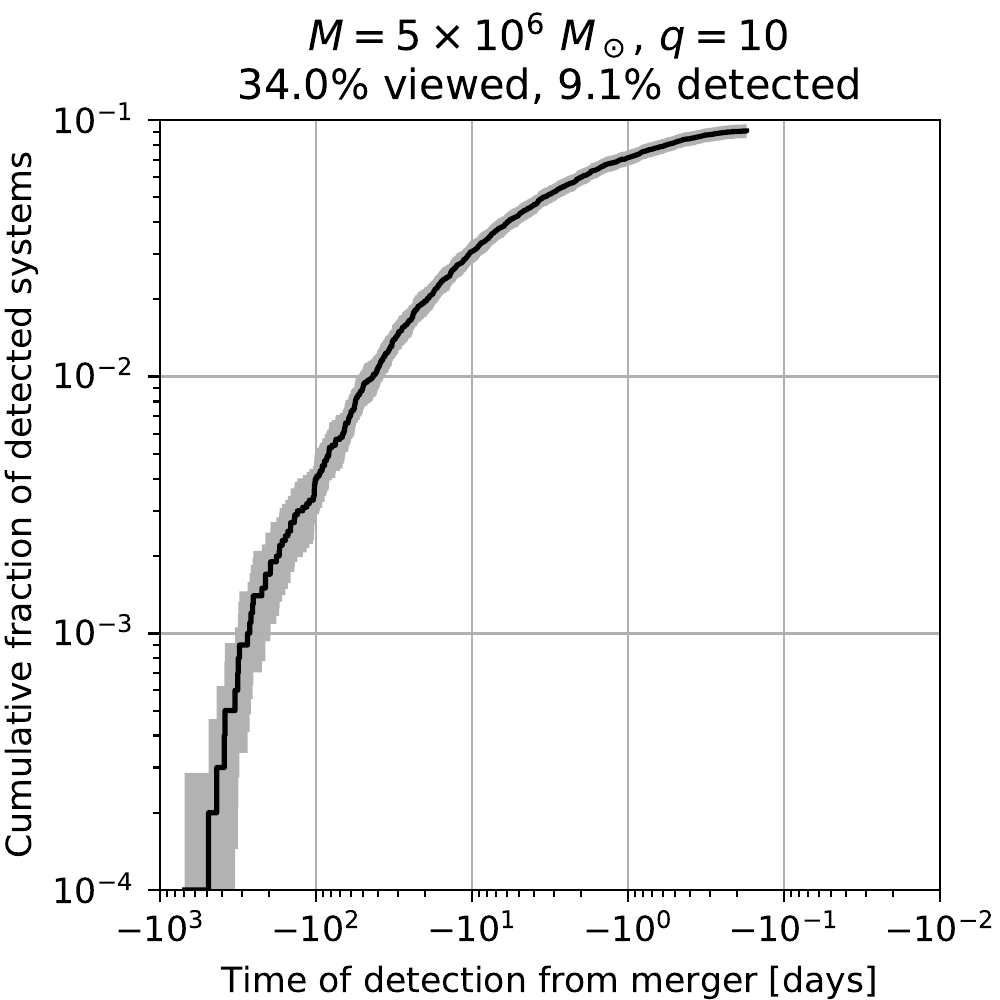}
    \caption{Fraction of LISA detections up to $z=3.5$ for which the X-ray modulation is detected by TAP/XRT at the time shown in the x axis or earlier. XRT exposures are fixed to $10^4$ s. The three plots correspond to different mass parameters, indicated above the plots. The shaded bands indicate the 95\% confidence intervals on the probability of success given the number of simulated and detected systems. About 1\% of the sources are detected at least months before merger, irrespective of their mass. Asymmetric systems produce the largest fraction of detections overall.}
    \label{fig:fraction_over_time_1}
\end{figure*}

A useful metric that can be extracted from these simulations is the fraction of systems detected at least a certain number of days before merger. This quantity is plotted in Fig.~\ref{fig:fraction_over_time_1} for the three populations considered above.
The lighter and heavier equal-mass populations produce similar overall detected fractions, with respectively $4\%$ and $5\%$ of the simulated systems resulting in an X-ray detection. The unequal-mass population leads to a higher detected fraction, approximately twice as large as the others. $1\%$ of the lighter systems produce a detection $\approx 100$ days before merger, while the heavy equal-mass systems are detected much later. Heavy unequal-mass systems fall in between the other two cases.

\subsection{Shortening the exposure}

In our baseline simulations we saw that a large fraction of the systems is missed because it is never viewed, especially for the heavier populations. We can thus ask whether reducing the exposure time may allow more systems to be viewed and eventually detected, the argument being that the target coverage of the LISA localization can be achieved in a shorter time, before the localization is updated.

\begin{figure*}
    \includegraphics[width=0.666\columnwidth]{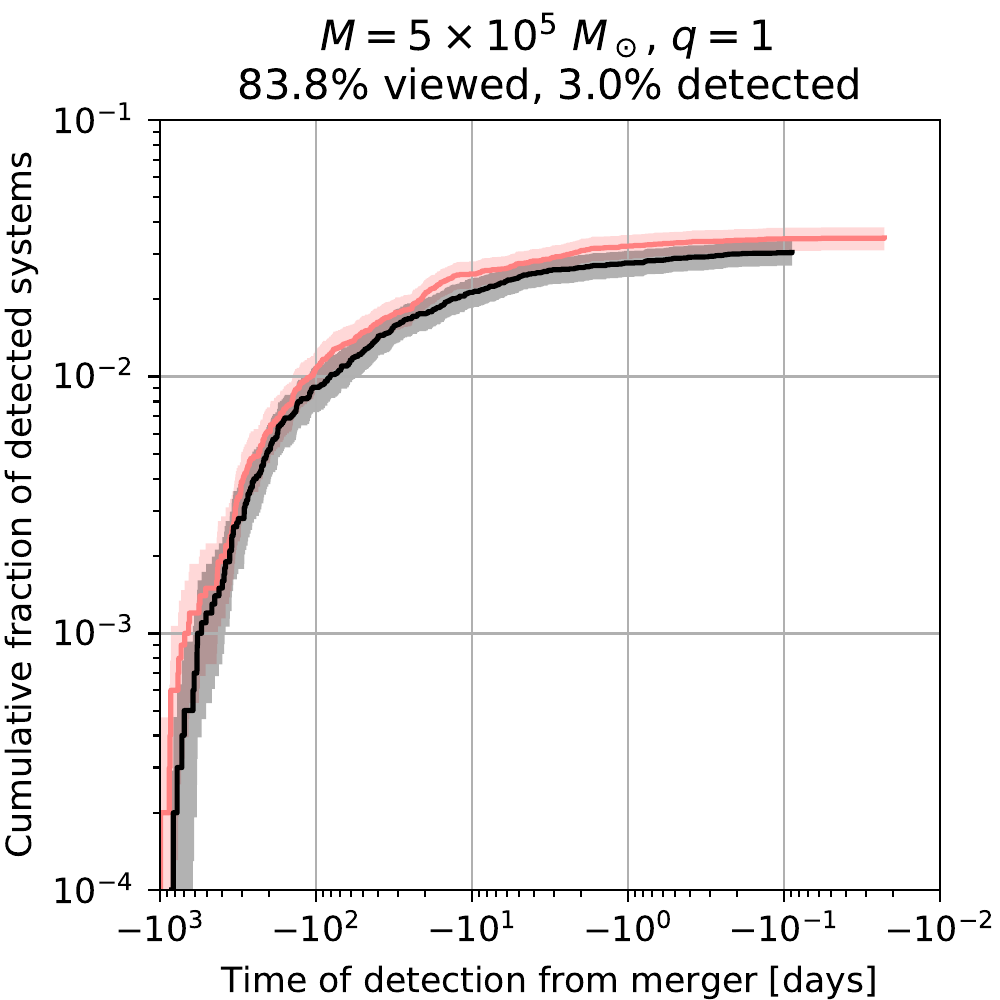}
    \includegraphics[width=0.666\columnwidth]{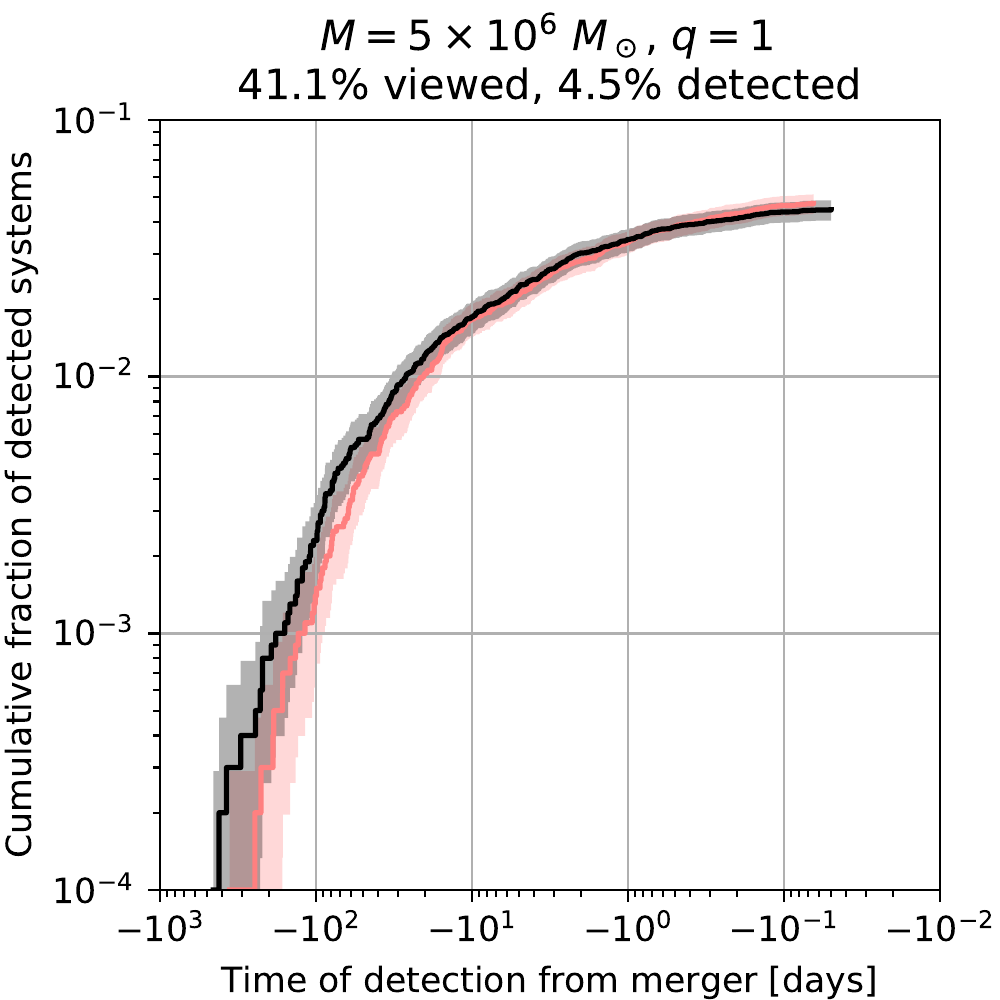}
    \includegraphics[width=0.666\columnwidth]{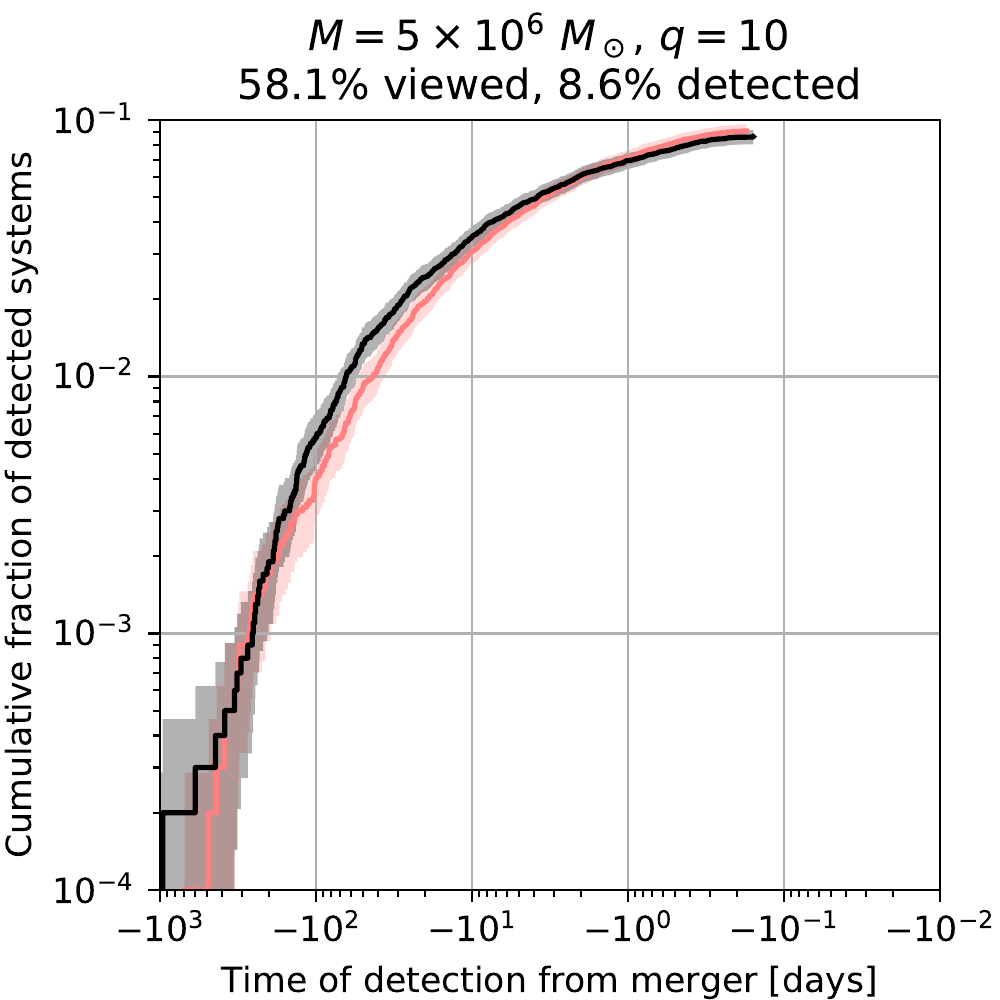}
    \caption{Fraction of LISA detections up to $z=3.5$ for which the X-ray modulation is detected by TAP/XRT, as a function of the detection time. The three plots correspond to different mass parameters, indicated above the plots. XRT exposures have been lowered to $10^3$ s. Although this increases the fractions of viewed systems, there is no net improvement with respect to a fixed exposure of $10^4$ s (lighter curves).}
    \label{fig:fraction_over_time_2}
\end{figure*}

To this end, we reduce the fixed exposure to $10^3$ s and repeat the simulations. The resulting fractions are plotted in Fig.~\ref{fig:fraction_over_time_2}.
As expected, we find that a larger fraction of the systems are now viewed by XRT at some point, almost a factor of 2 for the heavier populations with respect to the baseline. Nevertheless, reducing the exposure in this way does not appear to have a noticeable impact on the overall ability to detect the X-ray modulation: although more systems are viewed, the collected photon samples are not sufficient for the modulation to be detected.

Exposures shorter by another order of magnitude would become comparable with the time spent by TAP to slew across the large initial sky localizations. This would be a wasteful observing schedule, so we do not consider it for further exploration.

\subsection{Larger sky localization coverage}

\begin{figure*}
    \includegraphics[width=0.666\columnwidth]{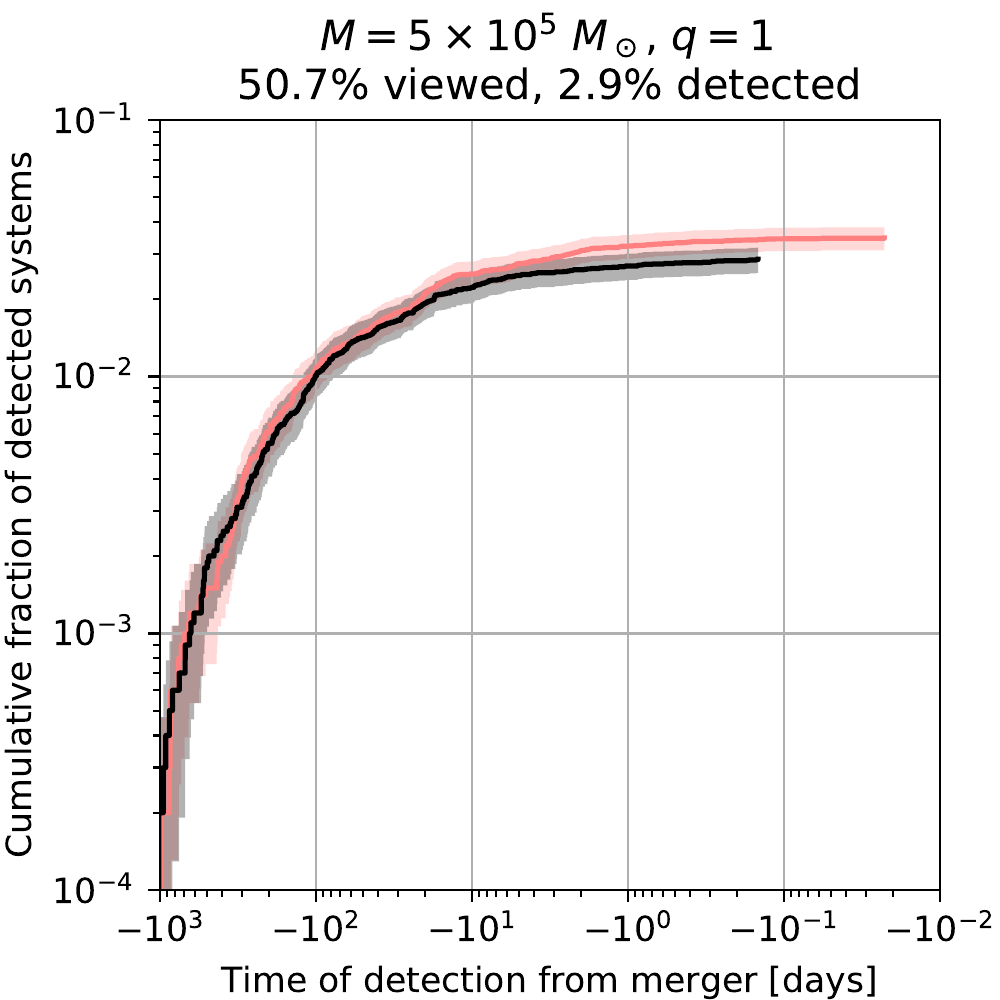}
    \includegraphics[width=0.666\columnwidth]{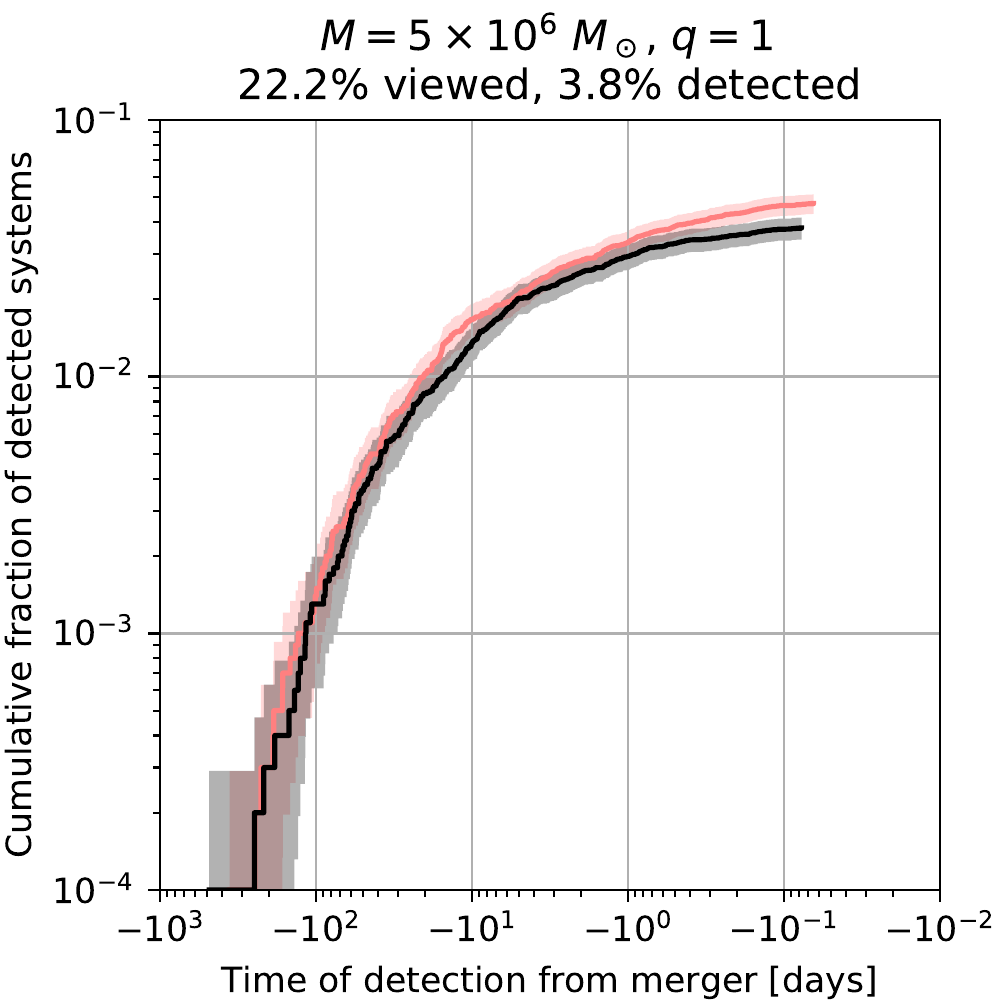}
    \includegraphics[width=0.666\columnwidth]{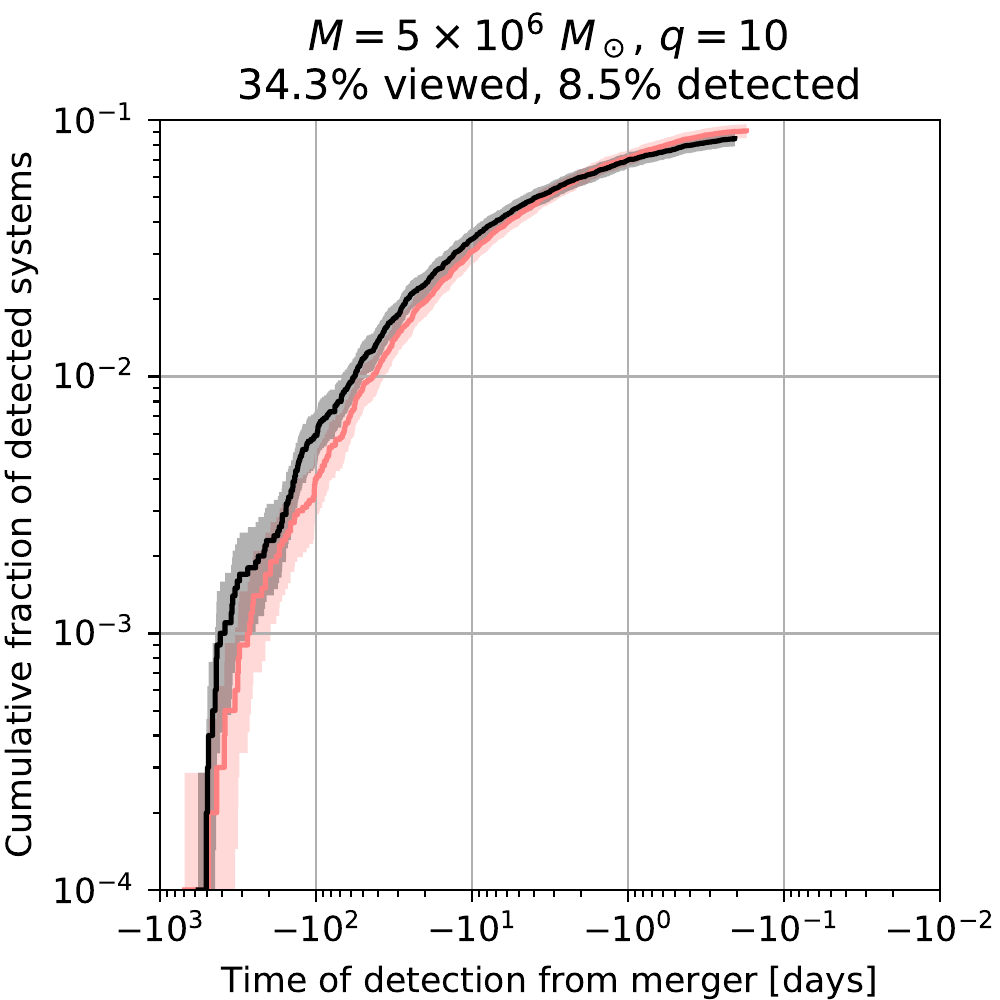}
    \caption{Fraction of LISA detections up to $z=3.5$ for which the X-ray modulation is detected by TAP/XRT, as a function of the detection time. The three plots correspond to different mass parameters, indicated above the plots. The target coverage of the LISA localization has been increased to 95\%, but the result is comparable to a 50\% coverage (lighter curves).}
    \label{fig:fraction_over_time_3}
\end{figure*}

The second variation we consider is raising the target sky localization coverage to 95\% and keeping the exposure time to the initial value of $10^4$ s. The resulting fractions are in Fig.~\ref{fig:fraction_over_time_3}.

The fractions of viewed and detected systems are still very close to the baseline. At an exposure of $10^4$ s only $\approx 17$ different sky locations can be viewed between updates, so this result suggests that the viewed fraction is actually limited by the exposure time: for many systems there is not enough time to complete even a 50\% coverage before LISA updates the skymap, so raising the coverage cannot increase the number of detections.

\subsection{Sky localization threshold}

Many of the simulated systems have essentially no sky localization shortly after reaching an SNR of 10 in LISA, and we have just seen that with our default exposure time we can only explore an area of $\approx 17$ deg$^2$ between each sky localization update. One may therefore argue that it is wasteful to start observing the systems so early, and rather wait for the sky localization to become sufficiently tight before starting the observation.

Here we repeat the simulations with the condition that observation starts as soon as the SNR is larger than 10 \emph{and} the $64\%$ credible region of the sky localization is smaller than 20 deg$^2$. This allows a large fraction of all sky localizations to be fully explored with exposures of $10^4$ s before receiving an update. It also frees up a large amount of TAP's observing time for other targets. On the other hand, for many binaries we will have much less time to detect the X-ray modulation before they merge.

\begin{figure*}
    \includegraphics[width=0.666\columnwidth]{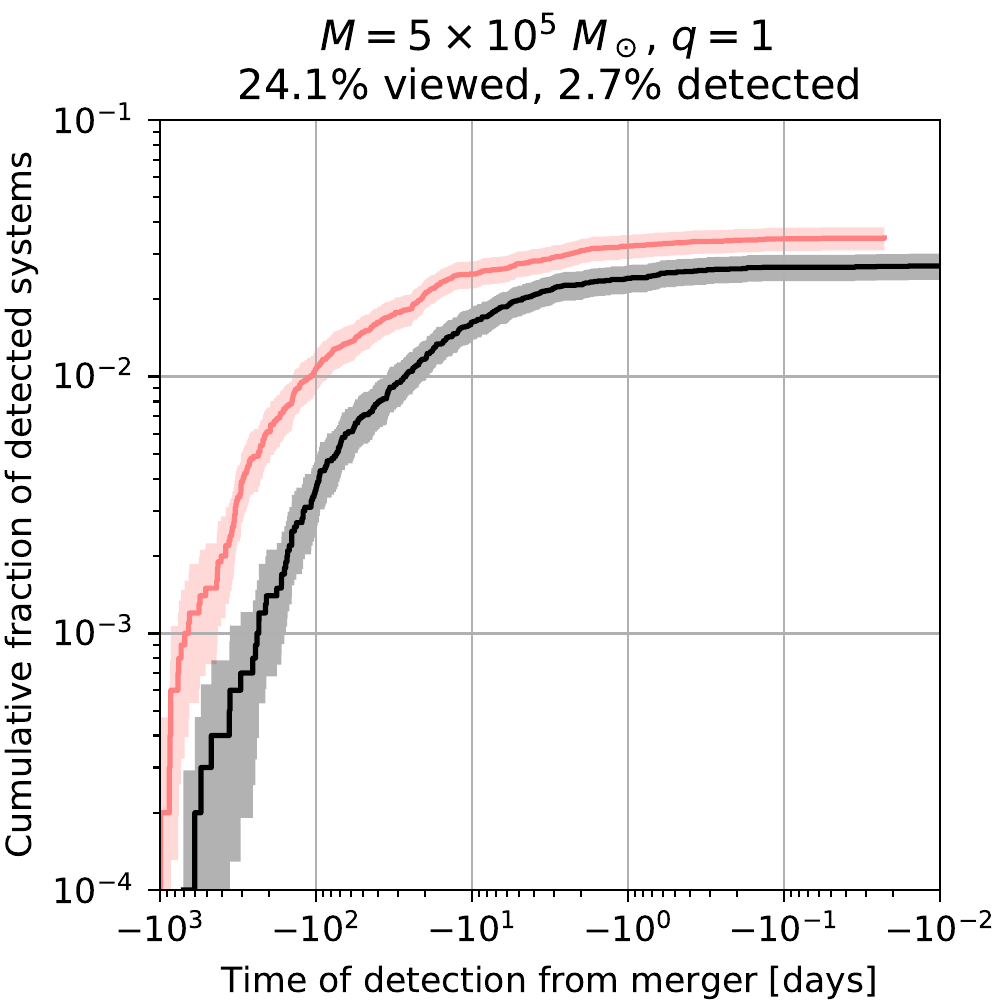}
    \includegraphics[width=0.666\columnwidth]{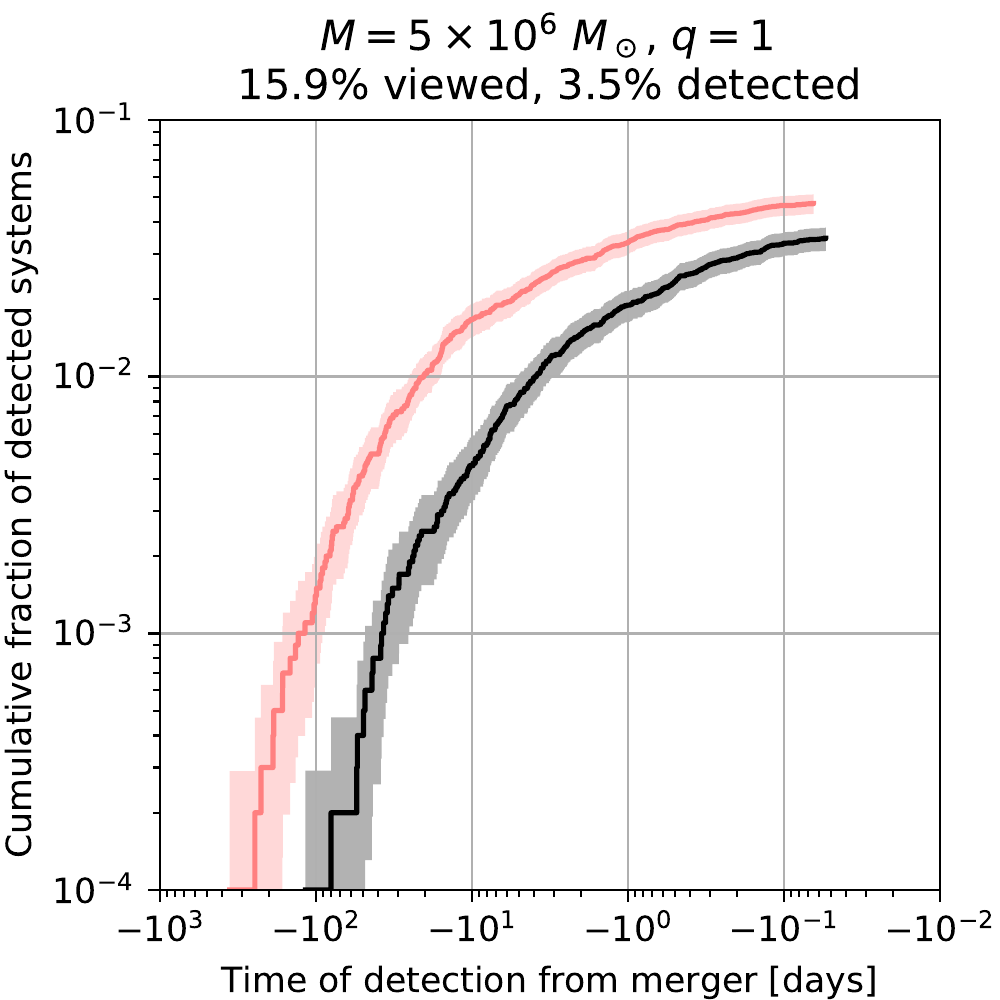}
    \includegraphics[width=0.666\columnwidth]{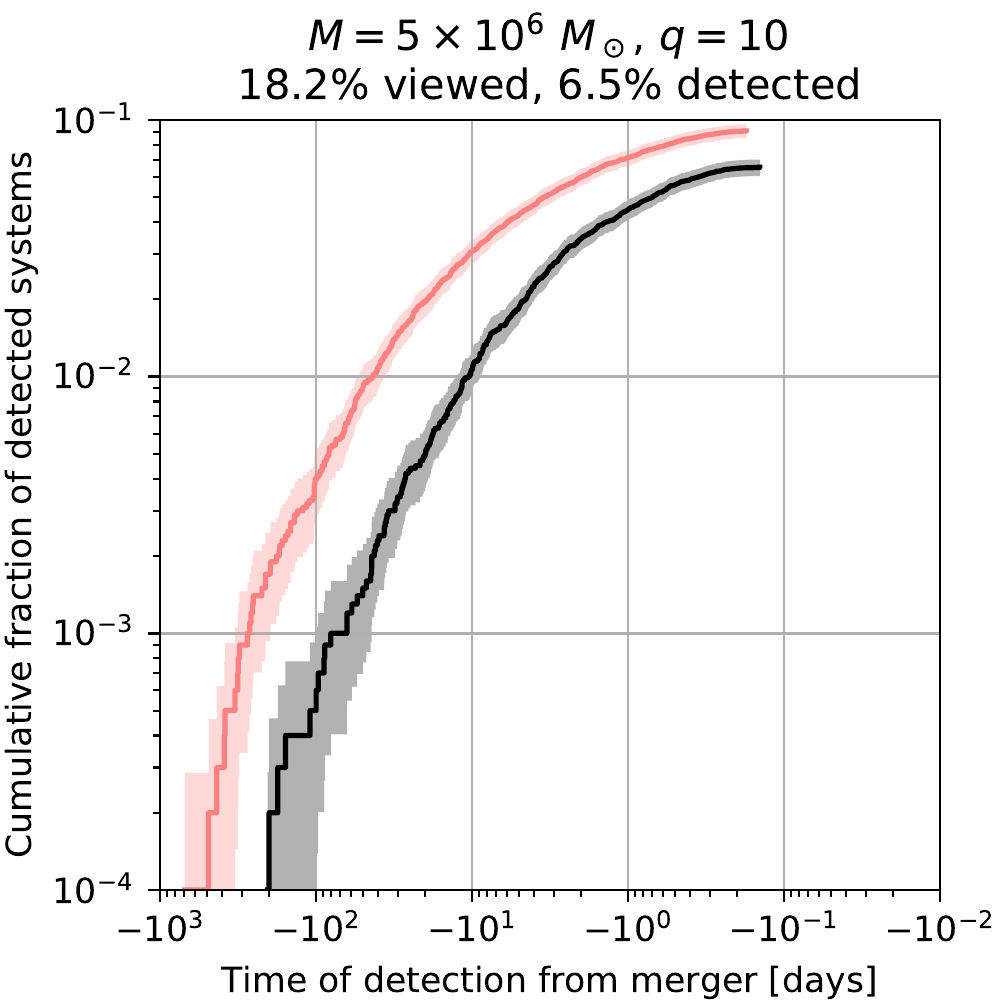}
    \caption{Fraction of LISA detections up to $z=3.5$ for which the X-ray modulation is detected by TAP/XRT, as a function of the detection time. The three plots correspond to different mass parameters. Here the LISA sky localization must be more precise than 20 deg$^2$ before starting the TAP/XRT observation. Lighter curves show the baseline results. The overall detection fractions are similar, especially for the lighter systems, but the detection times are pushed closer to merger.}
    \label{fig:fraction_over_time_area_threshold}
\end{figure*}

The result of the simulations is in Fig.~\ref{fig:fraction_over_time_area_threshold}.
We can see that the additional constraint reduces the overall detected fractions by less than $30\%$. The light symmetric population is the least affected, which is consistent with the fact that the high-frequency content of its gravitational-wave signals enables a precise localization earlier than heavier binaries. Regardless of the mass parameters, the detections are pushed noticeably closer to merger, but 1\% of the systems can still be detected at least a couple days before merger.

Based on these results, an optimal decision for triggering the observation might depend on SNR, sky localization area and masses of the binary, with heavier systems starting the observation earlier than lighter ones.

\subsection{Limited availability}
\label{sec:limavail}

The previous simulations have been using the idealistic assumption that an unlimited amount of TAP observing time is dedicated to the followup of each LISA detection. In practice this will not be feasible, not only due to the different scientific goals of the telescope, but also because some of the supermassive binaries detected by LISA might be inspiraling at the same time, thus requiring TAP to continuously hop between multiple systems.

Hence, here we take again the baseline configuration and introduce a ``dead time'' after each exposure to simulate the limited availability of the instrument. We set the dead time to 9 times the exposure time, i.e.~$9 \times 10^4$ s, implying that only $\approx 10\%$ of the TAP observing time is spent on each LISA target.

\begin{figure*}
    \includegraphics[width=0.666\columnwidth]{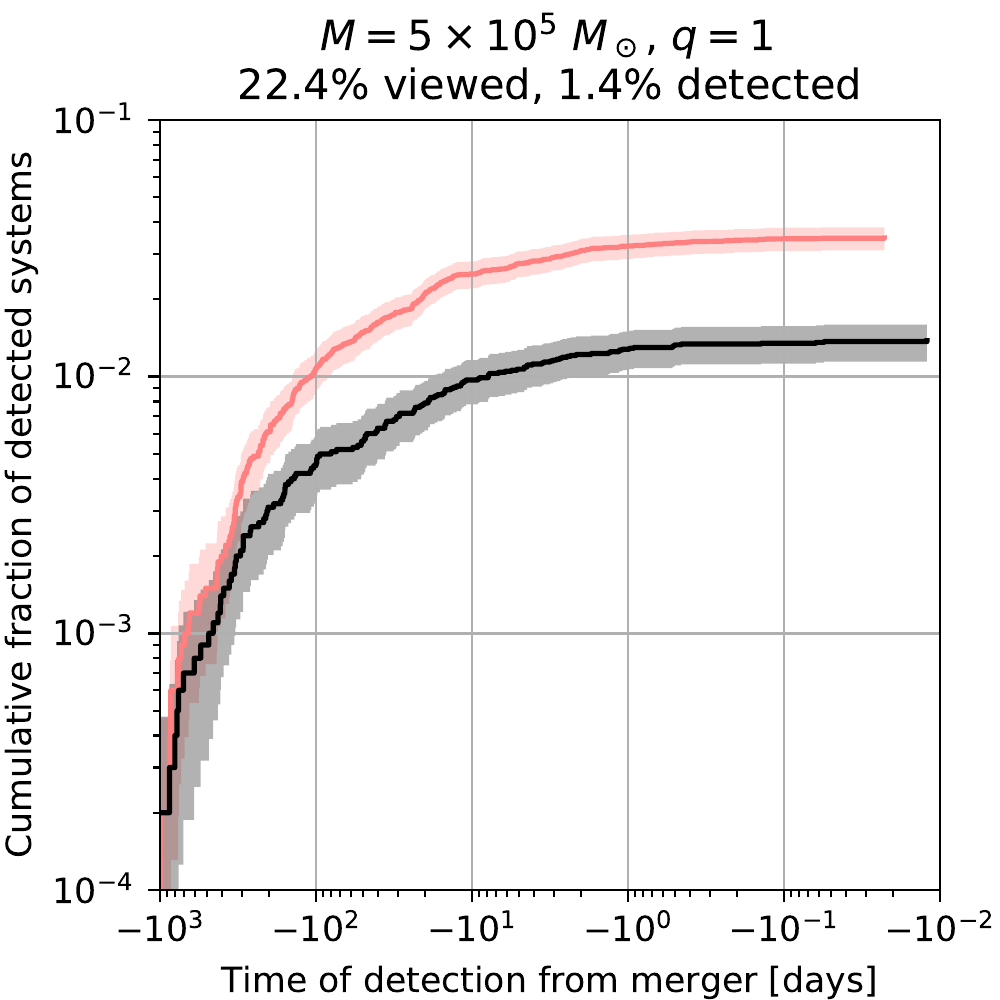}
    \includegraphics[width=0.666\columnwidth]{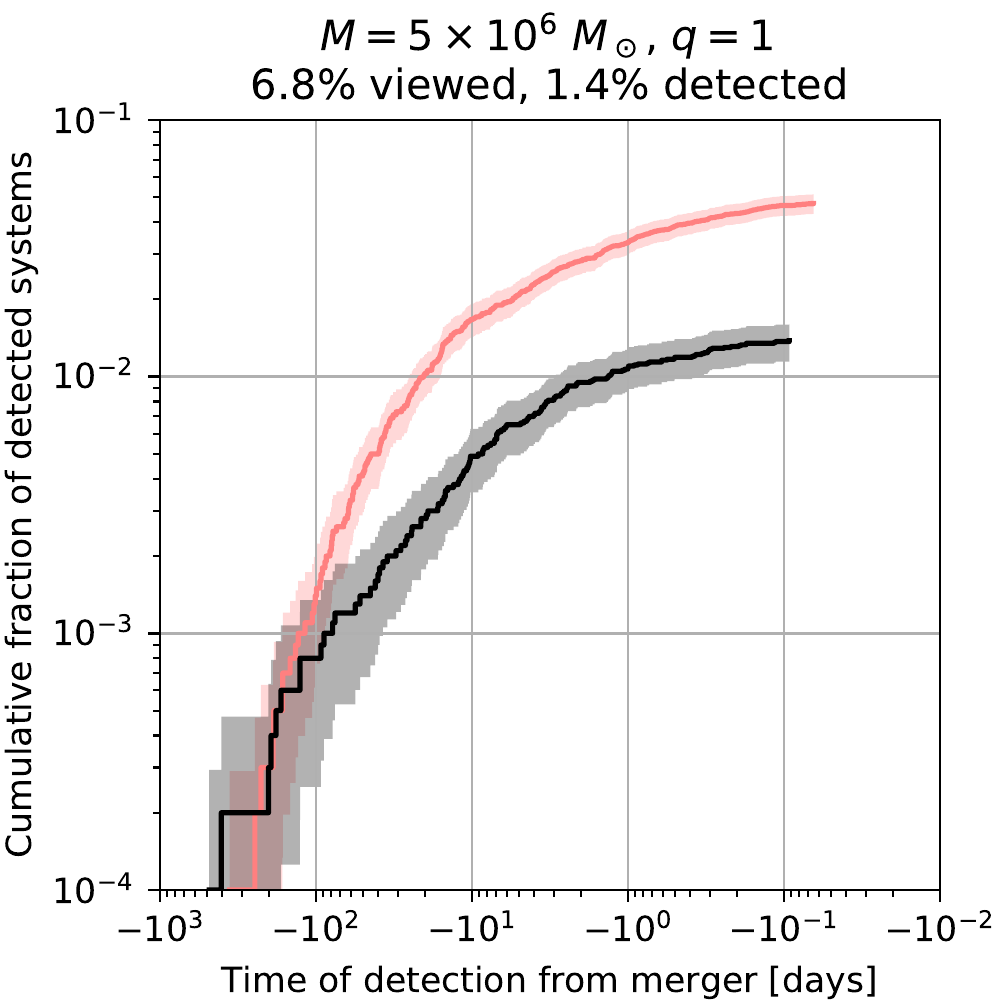}
    \includegraphics[width=0.666\columnwidth]{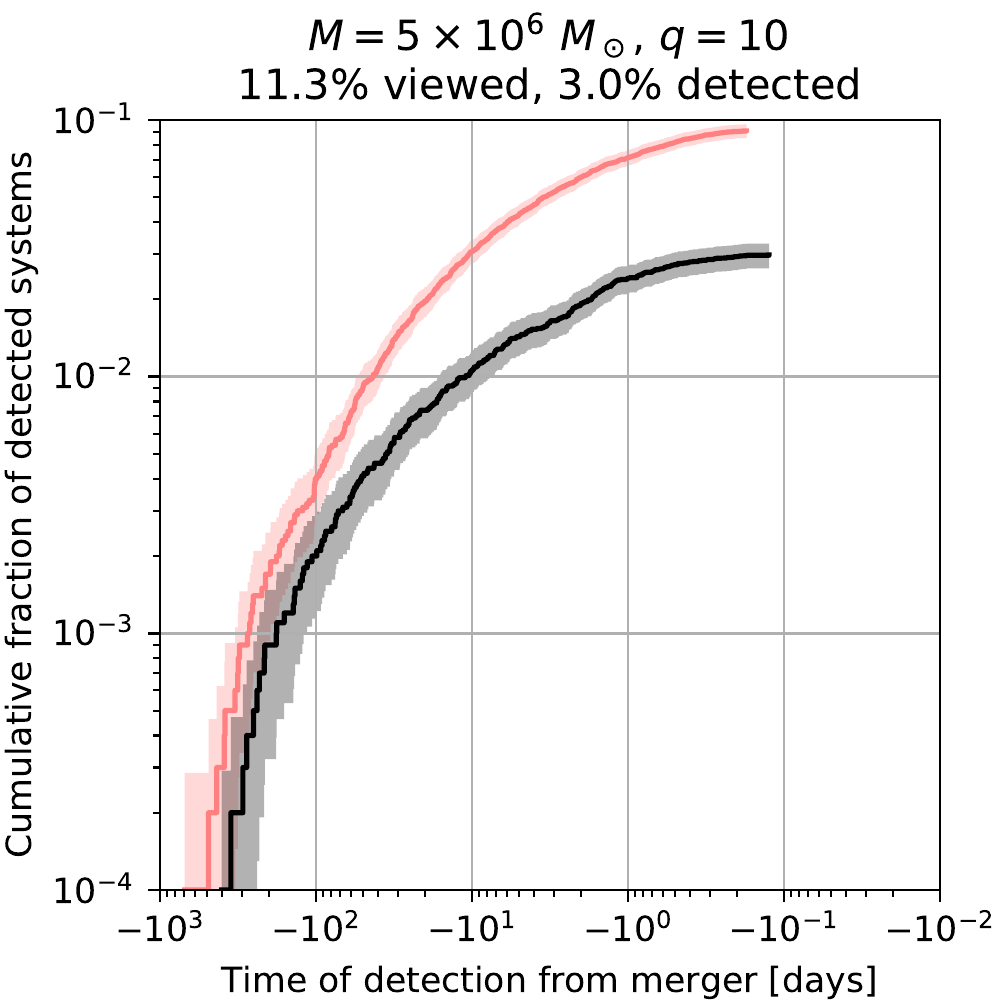}
    \caption{Fraction of LISA detections up to $z=3.5$ for which the X-ray modulation is detected by TAP/XRT, as a function of the detection time. The three plots correspond to different mass parameters. A $9 \times 10^4$ s dead time has been introduced after each XRT exposure, effectively reducing the TAP allocation for each source to 10\% only. Lighter curves show the baseline results. The fractions are visibly reduced and the detection times are pushed much closer to merger, but 90\% of the TAP observing time is now available to other targets.}
    \label{fig:fraction_over_time_wait}
\end{figure*}

The result is shown in Fig.~\ref{fig:fraction_over_time_wait}.
The overall number of detected systems is reduced by a factor of $\approx 3$, with a small dependence on the mass parameters. Nearby systems continue to be detected very early, despite the introduction of the dead time. Considering that we have freed 90\% of the spacecraft time for other science (or other supermassive black hole systems) this is a positive effective gain.

One can easily imagine more complicated observation schedules which would lie between our baseline and this fixed 10\% allocation, and would likely raise the detected fractions back towards our baseline results. For instance, once a particularly nearby binary is a few days away from its merger, it would become a high-priority target and its TAP allocation could be increased in an ``adaptive'' way. With the sets of simulations discussed above, we believe we have bracketed any realistic observing schedule, and the optimal strategy will be found in future studies.

\subsection{Detection rates and observing time}

So far we have discussed detection efficiencies, i.e.~fractions of LISA detections (up to $z = 3.5$) which also lead to a detection of the X-ray modulation. In this section we convert such fractions to expected detection rates, under the assumption that TAP will only focus on systems with redshift smaller than $3.5$. We also explore the amount of TAP observing time required by the different strategies presented earlier.

In order to do so, we turn again to the merger rate models shown in Fig.~3 of \citet{Klein:2015hvg}. In that plot, differential merger rates are given as a function of redshift for various assumptions about the population and evolution of supermassive black hole binaries. We consider the ``popIII'' and ``q3-nod'' models, but not the ``q3-d'' model, which is presented as a conservative case in that study. Integrating both models up to the maximum redshift used in our simulations ($z = 3.5$) gives comparable merger rates of order $10$ \peryear. Because of the uncertainties and model-dependency of these estimates, here we adopt a fixed fiducial merger rate of $10$ \peryear. The merger rate can be taken as a proxy for the LISA detection rate, because essentially all binaries in our simulations are detected by LISA with SNR larger than 10 at some point.

\begin{figure*}
    \includegraphics[width=0.666\columnwidth]{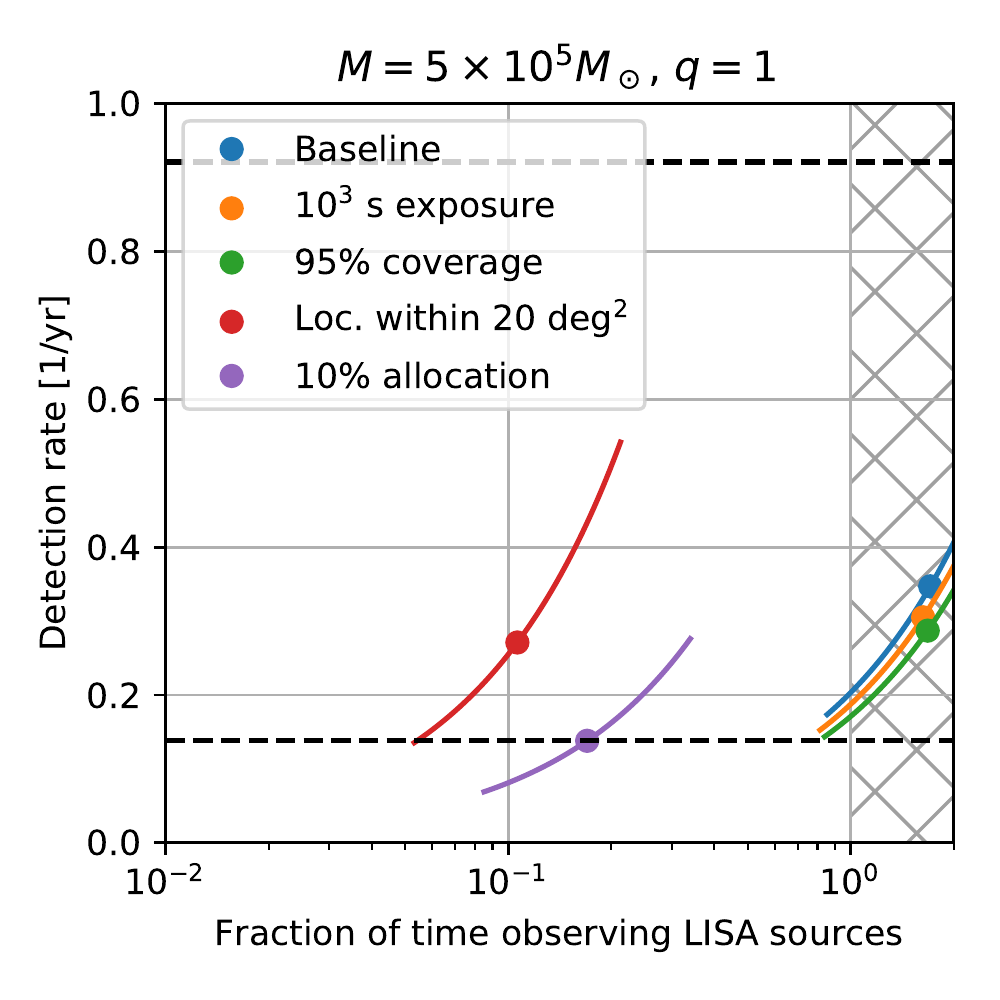}
    \includegraphics[width=0.666\columnwidth]{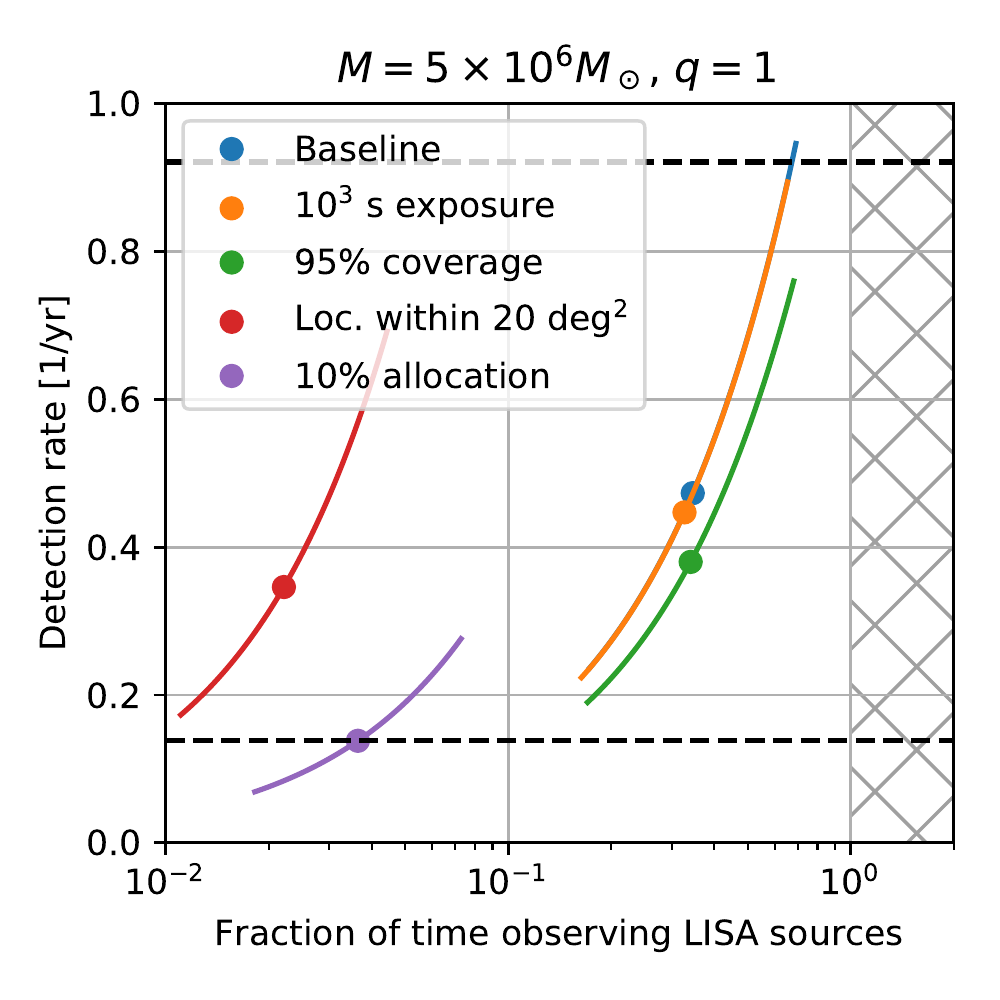}
    \includegraphics[width=0.666\columnwidth]{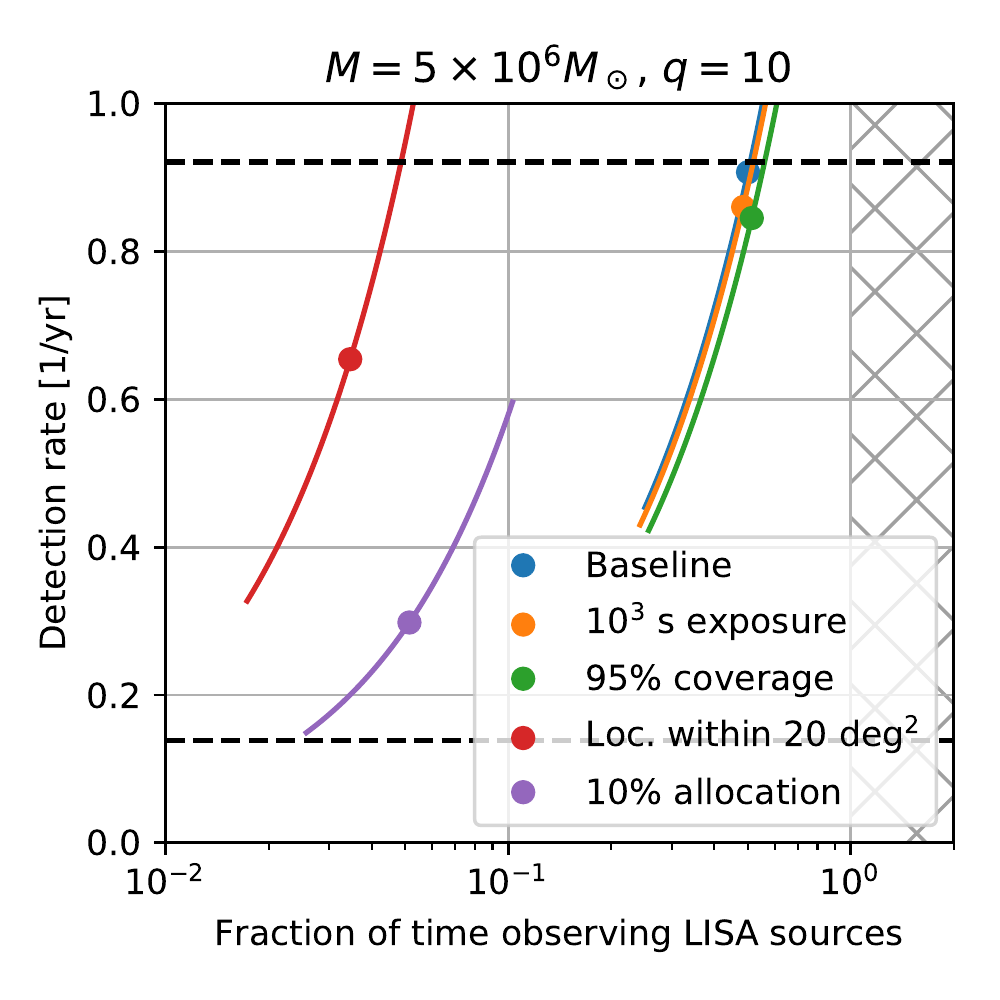}
    \caption{Comparison of the different observing strategies considered in our simulations, in terms of amount of required TAP observing time and resulting rate of detections, assuming a fiducial rate of 10 \peryear LISA detections at redshift less than $3.5$. The lines show what happens if the fiducial LISA rate is varied by a factor of 2 in both directions. The hatched areas indicate a regime where more than one TAP-like satellite would be required. The lower and upper dashed lines indicate the 50\% and 99\% probabilities of one or more detections in a 5 yr LISA mission. Although the baseline configuration leads to the highest detection rates, it also requires an unrealistic amount of TAP time, especially for lower-mass systems. Waiting for the LISA localization to become more precise than $20$ deg$^2$ leads to a much more reasonable balance between detection rate and observation time.}
    \label{fig:obstime}
\end{figure*}

Combining the LISA detection rate with the previous results, and with the fact that each simulation contains $10^4$ systems, we can calculate two figures of merit: (i) the rate at which we detect an X-ray modulation with XRT associated with a LISA detection, and (ii) the effective amount of time covered by each simulation. We then compute the total XRT observing time by simply summing the durations of all exposures performed in the simulations, and normalize the result by the amount of time covered by the simulations, which gives us the total fraction of TAP time spent observing LISA's supermassive black holes. We can then plot the detection rate as a function of the observing fraction for each of the simulated observing strategies. This plot is shown in Fig.~\ref{fig:obstime} and it includes the effect of varying the fiducial LISA rate by a factor of 2 in both directions.

We can see that our baseline configuration leads to the highest detection rates for all mass choices, at the price of requiring a large fraction of the TAP allocation. In fact, for systems with $M = 5 \times 10^5 \msun$, the fraction is even larger than 100\%, meaning that we would need \emph{more than one} TAP-like satellite to perform the observation. We find once again that lowering the exposure time or increasing the target coverage of the LISA localization probability does not produce significantly different results. On the other hand, waiting for the LISA localization to become sufficiently precise dramatically lowers the required TAP allocation, while reducing the detected rate by 30\% at most. A fixed 10\% allocation per-system also lowers the total allocation by a similar amount, although it reduces the detection rate by a factor of $2$ or $3$. We can thus see that the best approach among the ones we considered is to require a certain precision in the sky localization before starting the TAP observation, and then continue staring at the system until it has merged. A sky localization threshold larger than 20 deg$^2$ likely corresponds to a point between the ``baseline'' and ``loc.~within 20 deg$^2$'' points in Fig.~\ref{fig:obstime}. The same idea can be used to eyeball where a fixed per-system allocation between 10\% and 100\% would fit.

In this section we assumed that only systems at $z < 3.5$ are targeted by TAP. Based on Fig.~\ref{fig:time_inc_dist}, it appears that the redshift threshold could be lowered even more with a small impact on the detection rate and a (further) large reduction of the required observing time, especially for lower-mass systems. However, the ability to decide whether a system is sufficiently close to warrant starting an observation depends on how precise the luminosity distance measurement is at that point, and the tighter the restriction is on redshift or distance, the harder it will be to make a reliable decision. We expect that imposing a requirement on distance will affect the detected fractions in a way qualitatively similar to Fig.~\ref{fig:fraction_over_time_area_threshold}. The optimal decision to start the XRT observation will likely be based on the joint distribution of masses, luminosity distance, orbital inclination and sky location inferred from the LISA data. How to take this decision will be optimized in future studies, using more accurate LISA parameter inference procedures and more reliable X-ray lightcurve models.

\subsection{False positives from unrelated sources}

In addition to the true source, the LISA sky localization will likely include bright AGNs whose lightcurves will exhibit some degree of variability, typically with power spectra decaying like power laws (``red noise''). One might suspect that such unrelated sources, especially when observed for a relatively short time, could resemble a quasiperiodic signal and thus trick our followup strategy into producing a false positive and distracting other observers from the true source. Because of the chirpy signature of the gravitational wave phase template, in our previous simulations we assumed that this would not be the case. Using the same tools described above, here we present an initial check of how well this assumption holds.

We repeat our baseline simulations after replacing each source's Doppler signal with a red noise signal, i.e. Eq.~\ref{eq:xrayflux} is replaced by
\begin{equation}
    F = \frac{YLN}{4\pi d^2_L}.
\end{equation}
Here identical symbols have the same meaning as in Eq.~\ref{eq:xrayflux}; $Y=0.1$ is the same bolometric correction of Eq.~\ref{eq:xbolcor}, without the $a^{-1}$ term; and the stochastic red noise contribution modulates the flux through the factor
\begin{equation}
    N = 1 + \frac{A}{2} \tanh \left( \frac{n}{2} \right)
\end{equation}
with $0 \le A \le 1$ being a parameter that controls how strong the stochastic modulation is, and $n$ being a Gaussian random time series with a $f^{-1}$ amplitude spectrum, normalized to zero mean and unit variance. The hyperbolic tangent guarantees the positivity of $N$ in a smooth way, but does not represent any particular physical effect. The resulting flux still depends on the mass of the binary because of its Eddington luminosity, but it is otherwise completely unrelated to the gravitational wave signal, providing a proxy for a variable X-ray source unrelated to the LISA detection.

With this setup, we first check the rate of false positives using $A = 0$, which produces time-independent lightcurves and exactly corresponds to the null hypothesis of Kuiper's test. Out of $10^4$ sources, performed using the three mass choices assumed in the previous simulations, we do not observe any detection. Considering that our detection threshold is set to a $p$-value of 0.003, we would expect a few false alarms. This result is hence consistent with our previous note about the trials factor in Eq.~\ref{eq:ptrials} being overestimated.

When setting instead $A = 1$, we obtain false positive probabilities of 0.002, 0.012 and 0.013 respectively out of $10^4$ systems. The false detections are all associated with the closest systems, up to a redshift of $\approx 0.5$. The lowest-mass result is still compatible with the expected rate of false positives given our detection threshold, but the other mass choices produce a rate a few times higher than expected. A possible explanation for this difference is the faster phase evolution of the lower-mass chirp templates, which we expect to average out the large low-frequency amplitude fluctuations of the red noise more effectively. Hence, a steeper red noise spectrum than we have assumed, or a spectral break before the frequency band of the modulation, might lead to a lower rate of false positives than we find. Indeed, many AGN variability power spectra have a power-law break at frequencies below our band ($10^{-5}$ Hz and $10^{-3}$ Hz from an SNR 10 to merger), beyond which the spectrum is significantly steeper than $f^{-1}$ \citep{Markowitz2003}. On the other hand, we find that 50\% of the systems produce a false detection at an average flux between $10^{-14}$ and $10^{-13}$ erg s$^{-1}$ cm$^{-2}$ (note that according to our model, larger fluxes also have larger absolute variability) and this result does not appear to depend strongly on the mass parameters. Hence, the higher rate of false positives associated with heavier systems might just be due to the higher Eddington luminosity. It is also important to note that we assumed all red noise sources to be located within the final LISA sky localization. Most confusing sources will be located within the initial localization but outside the final localization, and will be ruled out during the joint observation.

Nevertheless, in light of this initial check, it may eventually be necessary to use a more robust statistic for dealing with bright sources of red noise than the naive application of Kuiper's test. For instance, \cite{Vaughan2010} proposed a statistic that explicitly compares a model consisting of pure red noise against a model consisting of red noise plus a periodic signal. We expect that increasing the robustness to red noise will come at the price of a generally lower rate of detections, detections being pushed closer in time to merger, or both effects, by an amount dependent on the distribution of brightness and variability of unrelated sources in the LISA sky localizations.

\section{Conclusion}

We have simulated populations of supermassive black hole mergers and some of the expected properties of their X-ray emissions, focusing on the quasiperiodic Doppler modulation due to the orbital motion. For each system, we simulated the information that LISA would give us from the system's gravitational-wave emission. We focused on the information relevant for planning and starting electromagnetic observations of the sources, i.e.~the time of first detection and the inferred sky location, which is periodically updated as more data is acquired by LISA. We then simulated a plausible campaign of electromagnetic followup of the LISA detections, using the parameters of the XRT instrument on the proposed TAP mission. From the results of the simulations, we derived expected rates of multimessenger detections with LISA and TAP/XRT. We considered different choices for some of the parameters determining how the followup is performed and studied how they affect the detection rates and the fraction of TAP time that would be required by this program.

Assuming the LISA mission will last 5 years and discover $\approx 10$ supermassive black hole mergers per year at $z < 3.5$, we find that detecting a Doppler-modulated X-ray signal is practically feasible with an instrument like TAP/XRT. Such a detection will most likely come from a relatively massive and asymmetric binary with an intermediate orbital inclination. With some luck, one of the detections might happen tens of days before merger, enabling a long-term simultaneous multimessenger observation of the chirping system via gravitational and electromagnetic waves and thus providing precision data about the dynamics of the accretion and the evolution of the accretion disk.

In order to boost the chance of a detection, a significant amount of TAP observing time should be dedicated to a binary at least in the final days of its inspiral. Starting to observe systems localized to hundreds of square degrees is wasteful: comparable detection rates, with a greatly reduced amount of dedicated time, can be obtained by waiting for the localization to become better than a few tens of square degrees. However, a delayed followup would significantly hinder the chance of making very early detections. We explored a few straightforward observational strategies and we find that the choices of XRT exposure and covered amount of LISA localization probability are not critical. Different strategies, if properly optimized, might lead to a higher rate of detections or smaller TAP observing fraction. For instance, the exposure could be made a function of the size of the sky localization, or the observations could be synchronized to the phase of the modulation in an attempt to cover the full phase cycle as soon as possible. Such avenues need to be explored in future studies.

The simulations presented here involve a number of necessary simplifications, described next, which could be lifted in future studies to make the results more trustworthy.

The presence of red noise in the X-ray lightcurves will lower the chance of a detection to some extent, by fuzzing the Doppler modulation of the true source and by increasing the chance of a false detection from an unrelated system within the LISA sky localization. We performed an initial study of false detections using a simple model for the unrelated sources, assuming both time-independent and red-noise fluxes. Time-independent sources do not appear to represent a problem, which is consistent with the assumptions behind Kuiper's test. On the other hand, bright and highly variable red-noise sources can produce a higher rate of false positives than expected. Hence, depending on the distribution of brightness and variability of these sources, a straightforward application of Kuiper's test might not be sufficient. This problem could be addressed in different ways. For instance, many of the brightest sources might simply be well characterized and understood at the beginning of the LISA mission, to the extent that they would no longer pose the risk of confusing the search. One could also consider augmenting Kuiper's test with signal-based vetoes or using a statistic inspired by \cite{Vaughan2010}, which may have a negative impact on the rate of detections. Future studies should investigate such possibilities and also simulate the effect of red noise associated with the true source.

Apart from red noise, actual X-ray lightcurves from accreting binaries are generally going to be more complicated than our simple-minded analytic model, likely including additional modulation effects on top of Doppler boosting, such as accretion variability, and possibly spectral dependence on time. Our model will ultimately need to be calibrated to lightcurves extracted directly from GRMHD simulations of massive black hole binaries, which will likely become available in the future. Modulations unrelated to Doppler boosting might be relatively unaffected by the orbital inclination and thus have a stronger effect than Doppler boosting alone \citep{Kelley:2018fur}. These effects would raise the rate of detections.

Our simulations assumed that every supermassive black hole merger produces an AGN-like emission, which can be argued considering that major galaxy mergers appear to be associated with the most luminous AGNs \citep{Treister2012}. In addition, \cite{Weigel2018} recently showed that the probability of a galaxy hosting an AGN is an order of magnitude greater in a merger galaxy than in the field. Combined with the observations of \cite{Wang2017}, this corresponds to a total AGN fraction exceeding 50\% in galaxies that have recently merged. An efficiency smaller than our assumed 100\% would obviously lower our expected detection rates by the same factor.

During the advanced merger stages, the black holes might be obscured by the surrounding gas and dust, as found in e.g.~\citet{Koss2018} and thus produce a dimmer emission in soft X-rays, lowering the detection rate. A reasonable proxy for the unobscured fraction is the Type 1 / Type 2 AGN fraction ($\sim 25\%$). This fraction increases with luminosity and redshift and may therefore be even larger for the systems more likely to produce a detection \citep{Suh2019}. Hence, we do not expect obscuration to greatly reduce our estimates.

It is possible that the early detection of a merger in LISA data will trigger a more frequent downlink of the data than we have assumed. If the downlinked data can be turned into an updated sky localization on a time scale of a few hours, it might increase the number of detections in the last couple of days before merger.

Finally, realistic LISA localization posterior distributions will be more complicated than the Fisher ellipses we have assumed, in particular they could be composed by several widely-separated modes. A completely correct modeling of LISA sky localizations requires Bayesian parameter estimation on order $10^4$ realizations of simulated LISA data. The locations of the Sun, Earth and Moon can then be made consistent with the LISA localization, and thus the source occultation by such objects can be accurately simulated. However, we expect this ``ultimate'' approach to be orders of magnitude more expensive than our study in terms of computational cost, and the required technology is currently under active research \citep{Marsat:2018oam}.

\acknowledgments

We thank John Baker, Sylvain Marsat and Neil Cornish for discussion regarding the LISA sky localization, Will Zhang for discussion about XRT and Judy Racusin for discussion about observing schedules.

TD was and SN is supported by an appointment to the NASA Postdoctoral Program at the Goddard Space Flight Center, administered by Universities Space Research Association under contract with NASA. TD thanks Nelson Christensen and the Artemis lab at the Observatoire de la Côte d'Azur for the kind hospitality during part of this work. AM acknowledges partial financial support from the INFN TEONGRAV specific initiative. AS is supported by the Royal Society.

We are grateful for computational resources provided by the LIGO Laboratory and supported by National Science Foundation Grants PHY-0757058 and PHY-0823459. This study made use of NumPy \citep{numpy}, SciPy \citep{SciPy}, Matplotlib \citep{Hunter:2007} and Astropy \citep{2013A&A...558A..33A,2018AJ....156..123A}.

\bibliography{references}{}

\end{document}